\documentclass[aps,prd,amssymb,amsmath,amsfonts,superscriptaddress,nofootinbib,reprint,showpacs, longbibliography]{revtex4-1}

\usepackage{tabularx}
\usepackage{graphicx}
\usepackage{lmodern}
\usepackage{amsmath,amssymb}
\usepackage{mathrsfs}
\usepackage{amsfonts}
\usepackage[utf8]{inputenc}
\usepackage{url}
\usepackage{gensymb}
\usepackage[colorlinks]{hyperref}
\usepackage{xcolor}
\usepackage{multirow}
\usepackage[normalem]{ulem}
\usepackage{orcidlink}
\usepackage{placeins}
\usepackage{mathtools}
\usepackage{colortbl}
\usepackage{booktabs,cellspace}
\usepackage{rotating}
\usepackage{float}
\usepackage{afterpage}

\def\apj{\rm{Astrophys.~J.}}
\def\apjl{\rm{Astrophys.~J.~ Lett.}}

\def\mnras{\rm{Mon.~Not.~R.~Astron.~Soc.}}
\def\prd{\rm{Phys.~Rev.~D}}

\def\ssr{\rm{Space~Sci.~Rev.}}
\def\nat{\rm{Nature}}

\definecolor{dodgerblue}{HTML}{1E90FF}
\definecolor{viennared}{HTML}{DA0A14}
\definecolor{ctorange}{HTML}{FF6C0C}
\definecolor{wales}{HTML}{ff0038}
\definecolor{benettongreen}{HTML}{009421}
\definecolor{ferrarired}{HTML}{ff2800}
\definecolor{austriawienpurple}{HTML}{441678}
\definecolor{hullcityamber}{HTML}{f5971d}
\definecolor{steiermarkgruen}{HTML}{006747}

\hypersetup{
     colorlinks=true,
     linkcolor=dodgerblue,
     filecolor=dodgerblue,
     citecolor = dodgerblue,      
     urlcolor=dodgerblue,
     }

\newcommand{\Birmingham}{School of Physics and Astronomy, University of Birmingham, Edgbaston, Birmingham, B15 2TT, United Kingdom}
\newcommand{\BirminghamGW}{Institute for Gravitational Wave Astronomy, University of Birmingham, Edgbaston, Birmingham, B15 2TT, United Kingdom}

\newcommand{\btheta}{{\boldsymbol{\theta}}}
\newcommand{\channel}{{{j}}}
\newcommand{\Lorb}{{{\boldsymbol{L}}}}
\newcommand{\hLorb}{{ \hat{\boldsymbol{L}} }}
\newcommand{\OmegaL}{{{\boldsymbol{\Omega}_L}}}
\newcommand{\Jang}{{{\boldsymbol{J}}}}

\newcommand{\bchi}{{{\boldsymbol{\chi}}}}
\newcommand{\fp}{{{f^{\prime}}}}

\DeclarePairedDelimiterX{\norm}[1]{\lVert}{\rVert}{#1}

\begin{document}

\title{Precision tracking of massive black hole spin evolution with LISA}

\author{Geraint Pratten \orcidlink{0000-0003-4984-0775}}
\email{g.pratten@bham.ac.uk}
\affiliation{\Birmingham}
\affiliation{\BirminghamGW}

\author{Patricia Schmidt \orcidlink{0000-0000-0000-0000}}
\email{p.schmidt@bham.ac.uk}
\affiliation{\Birmingham}
\affiliation{\BirminghamGW}

\author{Hannah Middleton \orcidlink{0000-0001-5532-3622}}
\email{hannahm@star.sr.bham.ac.uk}
\affiliation{\Birmingham}
\affiliation{\BirminghamGW}

\author{Alberto Vecchio \orcidlink{0000-0002-6254-1617}}
\email{av@star.sr.bham.ac.uk}
\affiliation{\Birmingham}
\affiliation{\BirminghamGW}

\date{\today}

\begin{abstract}
The Laser Interferometer Space Antenna (LISA) will play a vital role in constraining the origin and evolution of massive black holes throughout the Universe. In this study we use a waveform model (IMRPhenomXPHM) that includes both precession and higher multipoles, and full Bayesian inference to explore the accuracy to which LISA can constrain the binary parameters. We demonstrate that LISA will be able to track the evolution of the spins -- magnitude and orientation -- to percent accuracy, providing crucial information on the dynamics and evolution of massive black hole binaries and the galactic environment in which the merger takes place. Such accurate spin-tracking further allows LISA to measure the recoil velocity of the remnant black hole to better than $100\,\mathrm{km}\,\mathrm{s}^{-1}$ (90\% credibility) and its direction to a few degrees, which provides additional important astrophysical information on the post-merger association. Using a systematic suite of binaries, we showcase that the component masses will be measurable at the sub-percent level, the sky area can be constrained to within $\Delta \Omega_{90} \approx 0.01 \, \rm{deg}^2$, and the binary redshift to less than $0.01$.  
\end{abstract}

\maketitle 

\section{Introduction}
\label{sec:Introduction}
The Laser Interferometer Space Antenna (LISA) will allow us to observe the mHz gravitational-wave (GW) spectrum. One of the key science goals of the mission is to trace the origin, evolution, and merger history of massive black hole (MBHs) throughout the Universe \cite{Klein:2015hvg,Sesana:2021jfh,LISA:2022yao}, where $M \sim 10^4 - 10^9 M_{\odot}$. These sources will be detected with large signal-to-noise ratios (SNR), allowing us to precisely measure the mass and spin angular momenta of the constituent black holes. 

The population properties of the massive black hole binaries (MBHBs) that will be detected by LISA depends on a wide range of possible evolutionary pathways. Observing and characterising this population will provide vital insight into black hole seed formation, galaxy assembly, and the formation and evolution of structure throughout the Universe \cite{Yu:2002sq,Tanaka:2008bv,Volonteri:2010wz,Klein:2015hvg,LISA:2022yao}. Whilst there are numerous open issues, there are several mechanisms that are of key importance but are still dominated by large theoretical and observational uncertainties. For example, we have a comparatively poor understanding of the mechanisms that generate the initial black hole seeds at high redshift ($z \gtrsim 15$) and the mechanisms that drive formation of the first protogalaxies. This, in turn, leads to a large uncertainty in the anticipated detection rates and the underlying astrophysical population properties \cite{Klein:2015hvg,Barausse:2020mdt,Sesana:2021jfh}. 

MBHs are known to reside in the centres of galaxies, e.g. \cite{Kormendy:1995mbh,Ghez:1998ph,Gebhardt:2000mbh,Ferrarese:2000mbh,Schodel:2003mbh,Ferrarese:2004qr,Ghez:2008ms,Genzel:2010zy,Kormendy:2013dxa,EventHorizonTelescope:2019dse}, and there is evidence that some galaxies contain a MBH, even at the early cosmological times probed at high redshift \cite{Banados:2014lia,Wu:2015nat,Banados:2017unc}. Coupled with the hierarchical assembly of galaxies through repeated mergers \cite{Fakhouri:2010mbh,Giersz:2015mbh,Oleary2021:mbh,Volonteri:2021nat}, it is anticipated that LISA will observe numerous MBHBs across a wide range of cosmic epochs. However, there is huge uncertainty on our understanding of the astrophysical mechanisms that produced the initial seeds from which they grew. Did they originate from the collapse of heavy population III stars forming in low-metallicity environments (light seeds) \cite{Madau:2001sc,Heger:2002by,Hirano:2013lba} or from the collapse of large proto-galactic disks (heavy seeds) \cite{Haehnelt:1993yy,Loeb:1994wv,Bromm:2002hb,Begelman:2006db,Lodato:2006mbh,Agarwal:2012qm}? How did these black holes (BHs) subsequently grow? Accretion is an inevitable process and plays a key role in the growth of MBHs \cite{Yu:2002sq,Volonteri:2006ma,Ananna:2018uec}. However, there are other mechanisms that can drive the growth of MBHs, such as repeated hierarchical mergers \cite{Volonteri:2002vz,Hopkins:2005fb,Sesana:2007sh,Tanaka:2008bv}. Observations of MBHBs by LISA will provide us with a unique opportunity to tackle these questions and to constrain the evolutionary pathways responsible for the MBHs observed today.

For a binary of MBHs to merge, efficient mechanisms that dissipate the angular momentum of the system are required. Consequently, our current understanding of massive BH evolution can be broadly divided into distinct phases that are characterized by the astrophysical mechanisms driving angular momentum loss in the binary. The initial phase is primarily driven by dynamical friction caused by the gravitational interaction between the BHs and the surrounding gas and stellar material. Within this phase, the BHs are driven from a separation of approximately $\sim \mathcal{O}(10^{-2} \rm{Mpc})$ to $\sim \mathcal{O}(1 \rm{pc})$ \cite{Frank:1976lcs}. Below these separations, the dominant mechanism is hardening~\cite{Frank:1976lcs, Lightman:1977lcs, Quinlan:1996vp}. Hardening refers to the interaction of fast-orbiting stellar objects and the MBHB, resulting in energy being removed from the system. If the galactic host is sufficiently gas rich, a high-density circumbinary disk can form, leading to additional hardening of the binary through viscous dissipation~\cite{Begelman:1980nat,Gould:1999ia,Escala:2004jh}. At scales below the hardening separation, e.g. \cite{Vasiliev:2013nha}, GW emission becomes the dominant mechanism driving the inspiral of the binary \cite{Peters:1963ux}. However, depending on the physical properties of the BHs, and the environment of their galactic hosts, the astrophysical mechanisms at play can be widely different. 

A key tracer of the astrophysical formation mechanism will be the spin magnitude and orientation of the progenitor MBHs  \cite{Volonteri:2004cf, Bogdanovic:2007hp, Berti:2008af, Sesana:2014bea, Klein:2015hvg, Fiacconi:2017owt, Sesana:2021jfh, Volonteri:2021nat}, though these can be extremely challenging to measure \cite{Reynolds:2019uxi}. The orientation of the spin, in particular, is considered a key observable to discriminate between astrophysical formation pathways. When the BH spins are not aligned with the orbital angular momentum, they induce relativistic spin \textit{precession}, a phenomena that drives the time evolution of the spins and the orbital plane, leaving characteristic imprints in the emitted GW signal, see Sec.~\ref{sec:Precession}. 

Of particular interest is the role of the BH spins during the phase of disc-driven migration. Notably, the spin orientations of the BHs by the time they reach the GW driven regime will be sensitive to whether the galactic environment is gas-rich or gas-poor. 

If the environment is gas-rich, the inspiral of the binary can lead to the formation of a cavity in the circumbinary disc with accretion onto the BHs leading to the formation of minidiscs around each BH \cite{Roedig:2011mdc,Farris:2013uqa,Farris:2014zjo,Colpi:2014mbh,Bowen:2017oot}. If the viscous minidiscs are misaligned with respect to the BH spin, then Lense-Thirring precession (see \cite{Lense:1918zz}) will cause the inner disc to warp and dynamically relax, eventually reaching a coplanar configuration. This phenomenon is known as the Bardeen-Petterson effect \cite{Bardeen:1975bpe,Kumar:1985bpe,Scheuer:1996bpe,Natarajan:1998xt,Armitage:2002uu,Bogdanovic:2007hp}. However, the efficiency of Lense-Thirring precession is limited to the Bardeen-Petterson radius, beyond which the outer disc may maintain its original misalignment. A consequence, first noted in \cite{Rees:1978dal}, is that the outer disc will drive the secular alignment of the BH spin with the total orbital angular momentum of the disc itself. This process occurs on a longer timescale than the warping of the minidisc but on a faster timescale than the growth of the BH, i.e. we can treat the mass and spin of the BH as being fixed \cite{King:1999aq,Gerosa:2015xya}. As such, BH spins that are preferentially aligned with the orbital angular momentum are thought to be a clear tracer for the Bardeen-Petterson effect which, in turn, would suggest a gas-rich galactic host \cite{ColemanMiller:2013jrk,Gerosa:2015tea,Steinle:2022jmc}. Conversely, MBH mergers in gas-poor environments are not expected to undergo any such alignment and we would expect the spin orientations to be generically oriented. If the spins are preferentially aligned, we are de-facto reducing the amount of precession in the binary relative to the generically oriented configurations. Spin precession is therefore a key observable in unraveling astrophysical formation pathways and will provide valuable insights into the environment of the galactic host.

A caveat to the discussion above can occur due to the torque on the disc induced by a companion BH. These torques can lead to the break-up of the disc and potentially result in BH spins that are misaligned, even in gas-rich environments \cite{Dogan:2018xla,Gerosa:2020xly,Nealon:2021akj,Steinle:2022jmc}. A consequence is that this could introduce degeneracies between the environment and BH spin orientations \cite{Steinle:2022jmc}. 

Massive binaries, such as those discussed above, will have large signal-to-noise ratios in LISA, even at high redshift. For these binaries, we anticipate precision constraints on the masses and spins of the constituent BHs. Early pioneering work, see \cite{Vecchio:2003tn}, demonstrated the importance of spin-precession in breaking parameter degeneracies and in improving the precision to which the binary parameters could be constrained. This work was subsequently expanded in a series of papers, e.g. \cite{Lang:2006bsg,Lang:2007ge,Stavridis:2009ys,Lang:2011je}, that explored the impact of higher-order PN corrections, especially higher-order spin-orbit and spin-spin terms. Two particularly important conclusions are that the modulations in the GW signal induced by spin-precession allow one to accurately measure the BH spins and that precession also breaks the degeneracy between the binary orientation and its position on the sky, which could have important implications for identifying potential EM counterparts, e.g. \cite{Tang:2018rfm,DAscoli:2018dbt,Mangiagli:2020rwz}. 

LISA is sensitive to the entire sky, with the sensitivity depending on the source location and polarization, e.g. \cite{Cutler:1997ta}. The detector orbits the Sun such that its barycentere varies on a timescale of approximately one year. For long duration signals, the detector's translational motion induces a Doppler shift in the observed GW signal, which is sensitive to the angular position of the source \cite{Cutler:1997ta}. LISA also precesses around the normal to the ecliptic, leading to amplitude and phase modulations in the observed signal due to time-dependent variations in the antenna response pattern. The modulations of $h_+$ and $h_{\times}$ depend on both the sky-location and orientation of the binary \cite{Cutler:1997ta}. Moreover, information on the angular position is encoded in the relative amplitudes and phases of the polarizations. Incorporating higher multipoles into the waveform model used to analyse the data will break parameter degeneracies, notably the degeneracy between the distance and the inclination. By accurately modelling higher multipoles, a more precise determination of the angular position and distance, e.g. see \cite{Marsat:2020rtl,Pratten:2022kug} for the context of MBHB in LISA. 

When the spin angular momenta $\boldsymbol{S}_i$ are misaligned with the orbital angular momentum $\boldsymbol{L}$, precession occurs in the orbital plane and the spin vectors \cite{Apostolatos:1994mx, Kidder:1995zr}. This results in characteristic amplitude and phase modulations in the gravitational-wave signal, leading to several important observations. For example, in simple precession, the relative orientations of $\boldsymbol{L}$ and $\boldsymbol{S}_i$ precess about the total angular momentum $\Jang$ \cite{Apostolatos:1994mx}, causing time-dependent changes in the binary's inclination $\iota$, the polarization angle $\psi$, and the BH spins. The precession cycle's timescale is longer than the orbital timescale but typically shorter than the LISA timescale, depending on the masses and spins of the black holes. As a consequence, precession-induced modulations enrich the gravitational-wave signal with valuable information about the binary and allow us to further break parameter degeneracies \cite{Cutler:1992tc, Vecchio:2003tn, Lang:2006bsg, Lang:2007ge, Chatziioannou:2014coa, Pratten:2020igi}. The importance of precession in this context was first highlighted in the pioneering work of \cite{Vecchio:2003tn} and subsequently refined in \cite{Lang:2006bsg, Lang:2007ge, Klein:2009gza, Stavridis:2009ys, Lang:2011je}. 

In this paper, we revisit the Bayesian parameter estimation of precessing MBHBs using a state-of-the-art inspiral-merger-ringdown (IMR) waveform models, IMRPhenomXPHM \cite{Pratten:2020ceb,Garcia-Quiros:2020qpx,Pratten:2020fqn}. We use two canonical massive BH binary configurations and systematically assess the ability of LISA to resolve the individual BH parameters, including their component masses and spin-angular momenta. Previous analyses have typically relied on either inspiral-only Fisher analyses, e.g. \cite{Vecchio:2003tn,Lang:2006bsg,Lang:2007ge,Stavridis:2009ys,Lang:2011je,Klein:2015hvg,Mangiagli:2020rwz,Klein:2022rbf}, or have been restricted to aligned-spins, e.g. \cite{Marsat:2020rtl,Toubiana:2020cqv,Mangiagli:2020rwz,Katz:2022yqe,Digman:2022igm,Littenberg:2023xpl}. 

We find broad agreement in the precision to which binary parameters are measured compared to the inspiral-only analyses, e.g. \cite{Vecchio:2003tn,Lang:2006bsg,Lang:2007ge, Klein:2015hvg}. We demonstrate the ability of LISA to accurately constrain the spin orientation hours before merger and can provide tight constraints on the the kick velocity and orientation of the remnant BH. 

\section{Lisa Data Analysis}
\subsection{Time-Delay Interferometry}
\label{sec:LisaDA}
Here, we broadly follow \cite{Marsat:2018oam,Marsat:2020rtl} to produce the LISA time-delay interferometry (TDI) response. In this framework, we apply a delay $d(t)$ to a signal $h(t)$ followed by a modulation $F(t)$ which, in the notation of \cite{Marsat:2018oam}, can be written as
\begin{align}
h_d (t) &= h(t + d(t)) , \\
s(t) &= F(t) \, h_d (t).
\end{align}
\newline 
The Fourier transform of the signal can then be written in terms of a transfer function $\mathcal{T}$ 
\begin{align}
\tilde{s} (f) &= \mathcal{T}(f) \, \tilde{h}(f) , \\
&= \int d \fp \, \tilde{h}(f - \fp) \, \tilde{G}(f - \fp, \fp) .
\end{align}
\newline 
Following \cite{Marsat:2018oam}, the transfer function can be decomposed into a set of signal-dependent timescales that correspond to different physical effects. The first timescale of interest is 
\begin{align}
T^2_f = \frac{1}{4 \pi^2} \left| \frac{d^2 \varphi}{d f^2} \right|,
\end{align}
which can be shown to reduce to the radiation-reaction timescale when the stationary phase approximation (SPA) holds \cite{Marsat:2020rtl}
\begin{align}
T^{\rm SPA}_f = \frac{1}{\sqrt{2 \dot{\omega} (t^{\rm SPA}_f) }},
\end{align}
where $\omega$ is the orbital phase of the binary. This naturally leads to a time-frequency correspondence of  
\begin{align}
t_{\ell m} (f) = t_c - \frac{1}{2 \pi} \frac{d \varphi_{\ell m} (f)}{d f},
\end{align} 
\newline 
where $t_c$ denotes the coalescence time of the binary. The effect of higher-order phase corrections to the transfer function can be written in terms of the time-frequency map as \cite{Klein:2014bua,Marsat:2018oam}
\begin{align}
\mathcal{T}_{\rm phase} (f) &= \displaystyle\sum_{p \geq 0} \frac{1}{p!} \left( \frac{i}{8 \pi^2} \frac{d^2 \varphi}{d f^2} \right)^p \left( \frac{\partial^{2p}}{\partial t^{2p}} G \right) (f,t_{\ell m} (f)),
\end{align}
\newline 
where $G$ denotes a frequency-dependent kernel \cite{Marsat:2018oam}. This expansion is related to the shifted uniform asymptotics expansion (SUA) introduced in \cite{Klein:2013qda, Klein:2014bua}. In this study, we focus on the leading order approximation, i.e. $p=1$, and defer a detailed analysis of the impact of higher-order corrections to the transfer function to future work.

The second key timescales arises from amplitude corrections beyond the leading order \cite{Marsat:2018oam}
\begin{align}
    \left( T_{A_p} \right)^p = \frac{1}{(2 \pi)^p} \frac{1}{A(f)} \left| \frac{d^p A}{d f^p} \right|,
\end{align}
\newline 
which leads to a transfer function of the form \cite{Marsat:2018oam}
\begin{align}
\mathcal{T}_{\rm amp} (f) &= \displaystyle\sum_{p \geq 0} \frac{1}{p!} \left( T_{A_p} \right)^p \left( \partial^p_t G \right) (f,t_f) .
\end{align}
\newline
Here we incorporate amplitude corrections up to second order, i.e. $p=2$. 

In practice, the TDI response is constructed in terms of a frequency-dependent transfer function acting on a set of waveform modes
\begin{align} 
h^{\rm A,E,T}_{\ell m} = \mathcal{T}^{\rm A,E,T} (f, t_{\ell m} (f)) \, h_{\ell m} (f).
\end{align} 
\newline 
The modes are generated using the precessing, higher-multipole inspiral-merger-ringdown model IMRPhenomXPHM \cite{Pratten:2020ceb,Pratten:2020fqn,Garcia-Quiros:2020qpx}. The term $\mathcal{T}^{\rm A,E,T} (f, t_{\ell m} (f))$ denotes the TDI transfer function and $\lbrace A, E, T \rbrace$ the 1G TDI channels, which are related to the original 1G Michelson TDI variables $\lbrace X, Y, Z \rbrace$ by \cite{Prince:2002hp,Tinto:2004wu}
\begin{align}
A &= \frac{1}{\sqrt{2}} (Z - X), \\
E &= \frac{1}{\sqrt{6}} ( X - 2 Y + Z), \\
T &= \frac{1}{\sqrt{3}} ( X + Y + Z).
\end{align}
\newline 
One of the key advantages of using the $\lbrace A, E, T \rbrace$ variables is that they form a set of noise-orthogonal variables \cite{Prince:2002hp}. As such, we assume that the noise in each channel is uncorrelated resulting in a diagonalized noise matrix that simplifies the resulting analysis \cite{Tinto:2004wu}. Note that the original variables introduced in \cite{Prince:2002hp} were constructed from the Sagnac variables $\lbrace \alpha, \beta, \gamma \rbrace$ and are therefore slightly different from the definition adopted here and in recent work, e.g. see the discussion in \cite{Buscicchio:2021dph,Klein:2022rbf}. Finally, we note that in order to generate the $\lbrace X, Y, Z \rbrace$ TDI variables, we adopt the rigid adiabatic approximation (RAA) \cite{Rubbo:2003ap,Marsat:2018oam}, in which 
\begin{align}
\tilde{X}_{1.5G} (f) \approx \left( 1 - e^{- 4 \pi i f L} \right) \tilde{X}_{\rm RAA} (f),
\end{align}
\newline 
where $L = 2.5 \times 10^9$m is the mean LISA armlength. 

\subsection{Bayesian Inference}
A central aim of Bayesian inference is to reconstruct the posterior distribution $p(\btheta | d)$ for the source parameters $\btheta$ given data $d$. From Bayes' theorem we have 
\begin{align}
p(\btheta | d) &= \frac{\mathcal{L}(d | \btheta) \pi(\btheta)}{\mathcal{Z}},
\end{align}
\newline 
where $\mathcal{L}$ denotes the likelihood of the data given the parameters $\btheta$, $\pi$ the prior distributions for $\btheta$, and $\mathcal{Z}$ the evidence. We perform Bayesian inference using a coherent analysis of the full LISA TDI output, $d = \lbrace d_{\channel} ; \channel = A, E, T \rbrace$, using the \texttt{Balrog} code. In particular, we employ nested sampling \cite{Skilling:2004nes,Skilling:2006nes} as implemented by \texttt{Dynesty} \cite{Speagle:2020dyn,Koposov:2023dyn}. The likelihood of the data $d$ given a set of binary parameters $\boldsymbol{\theta}$ is \cite{Cutler:1994fla} 
\begin{align}
    \label{eq:likelihood}
    \ln \mathcal{L} (d | \btheta) &= - \displaystyle\sum_{\channel} \frac{\langle d_{\channel} - h_{\channel} (\btheta) | d_{\channel}- h_{\channel} (\btheta) \rangle_{\channel} }{2} + \rm{const},
\end{align}
\newline 
where $h_{\channel}$ denotes the TDI output for channel $\channel$ produced by the GW signal. The noise weighted inner product is defined by 
\begin{align}
    \langle a | b \rangle_{\channel} &= 2 \int^{\infty}_0 df \, \frac{\tilde{a}(f) \tilde{b}^{\ast}(f) +  \tilde{a}^{\ast}(f) \tilde{b}(f) }{S_{\channel} (f)},
\end{align}
\newline 
where $\tilde{a}(f)$ denotes the Fourier transform of the time series $a(t)$ and $S_{\channel} (f)$ denotes the noise power spectral density (PSD) of the $\channel$-th TDI channel. We use the noise spectral densities provided by the ESA Science Requirements Document \cite{SciRD,lisa:2022psd} and include the unresolved galactic confusion noise according to the analytical fit of \cite{Babak:2017tow}. As in \cite{Pratten:2022kug}, we adopt a low-frequency cut-off of $f_{\rm low} = 10^{-4}$ Hz, corresponding to the pessimistic scenario in which LISA only retains sensitivity over the frequency window specified in the mission design requirements \cite{SciRD,lisa:2022psd}. Any sensitivity at lower frequencies will help improve the constraints on measured parameters, so our results can be interpreted as a lower bound on the LISA science performance. Finally, we adopt the zero-noise approximation such that $\tilde{n}(f) = 0$, noting that the noise still enters the likelihood in Eq.~\eqref{eq:likelihood} through the PSD. We also assume that the PSD is constant over the observation duration, though in reality we would expect the PSD to be non-stationary and to vary slowly in time, e.g. \cite{Edwards:2020tlp}. 

\section{Measuring Precession}
\label{sec:Precession}
\subsection{Black Hole Spins and Precession}
Spin-induced precession is a unique and exciting phenomena that arises in the general-relativistic two-body problem \cite{Barker:1970pre,Barker:1975ae,Apostolatos:1994mx,Kidder:1995zr,Damour:2001tu}. If the black hole spins are misaligned with the orbital angular momentum, $\Lorb$, spin-orbit and spin-spin couplings induce \textit{precession}. The spin-orbit couplings drive a time-dependent evolution of the orbital plane, i.e. the direction of $\Lorb$, whereas the relativistic spin-spin couplings, which first enter at 2PN, induce a torque on the spin-angular momenta that results in the \textit{nutation} of $\Lorb$, e.g. \cite{Apostolatos:1994mx,Kidder:1995zr}. The time evolution of $\Lorb$ can be decomposed into terms corresponding to precession and a term driven by radiation reaction 
\begin{align}
\frac{d \Lorb}{dt} &= \frac{d \hLorb}{dt} L + \frac{d L}{dt} \hLorb, \\
&= \underbrace{(\OmegaL \times \Lorb)}_{\substack{\text{precession}}} \, L + \underbrace{\frac{dL}{dt}}_{\substack{\text{radiation} \\ \text{reaction}}} \hLorb.
\end{align}
\newline 
Here $\OmegaL$ denotes the spin precession frequency in terms of a contribution arising from each of the BH spins
\begin{align}
\label{eq:sprecfreq}
    \OmegaL = \Omega_1 \boldsymbol{\chi}_1 + \Omega_2 \boldsymbol{\chi}_2, 
\end{align}
where the contribution to the precession frequency from each spin can be written at next-to-leading order in terms of the effective aligned spin \cite{Damour:2001tu,Racine:2008qv,Ajith:2009bn}
\begin{align}
    \chi_{\rm eff} = \frac{\bchi_1 \cdot \Lorb + q \, \bchi_2 \cdot \Lorb}{1 + q}
\end{align}
as \cite{Racine:2008qv,Gerosa:2020aiw}
\begin{align}
\label{eq:sprecfreq1}
    \Omega_1 &= \frac{{M^2}}{{2r^3(1+q)^2}} \left( 4 + 3q  - \frac{{3 q \chi_{\text{eff}}}}{{(1+q)}} \frac{M^2}{L} \right) , \\
    \Omega_2 &= \frac{{qM^2}}{{2r^3(1+q)^2}} \left(4q + 3 - \frac{{3q \chi_{\text{eff}} }}{{(1+q)}} \frac{M^2}{L} \right),
\label{eq:sprecfreq2}
\end{align}
\newline 
where $q = m_2 / m_1 \leq 1$ is the mass ratio of the binary.

The dynamical effects that govern quasi-circular precessing binaries evolve on three distinct timescales. The orbital timescale is governed by $t_{\rm orb} \propto (r/M)^{3/2}$, the precession timescale $t_{\rm prec} \propto (r/M)^{5/2}$ \cite{Apostolatos:1994mx,Kidder:1995zr}, and the radiation reaction timescale by $t_{\rm rad} \propto (r/M)^{4}$. At large separations, $r \gg M$, the timescales obey a hierarchy 
\begin{equation}
t_{\rm orb}  \ll t_{\rm prec} \ll t_{\rm rad}.
\end{equation}
\newline 
In the context of LISA, we expect MBH of total mass $\sim 10^4-10^7 M_{\odot}$ to be observable for $\mathcal{O}(\rm days)$ to $\mathcal{O}(\rm years)$, meaning that the binaries can undergo a comparatively large number of precession cycles compared to current ground-based observations of stellar mass black holes \cite{LIGOScientific:2018mvr,LIGOScientific:2020ibl,LIGOScientific:2021djp}. Therefore, we can anticipate that MBHB observations with LISA will provide valuable opportunities for precision measurements of black hole spins and a comprehensive understanding of their evolution over multiple precession cycles.

\begin{figure}
     \includegraphics[width=\columnwidth]{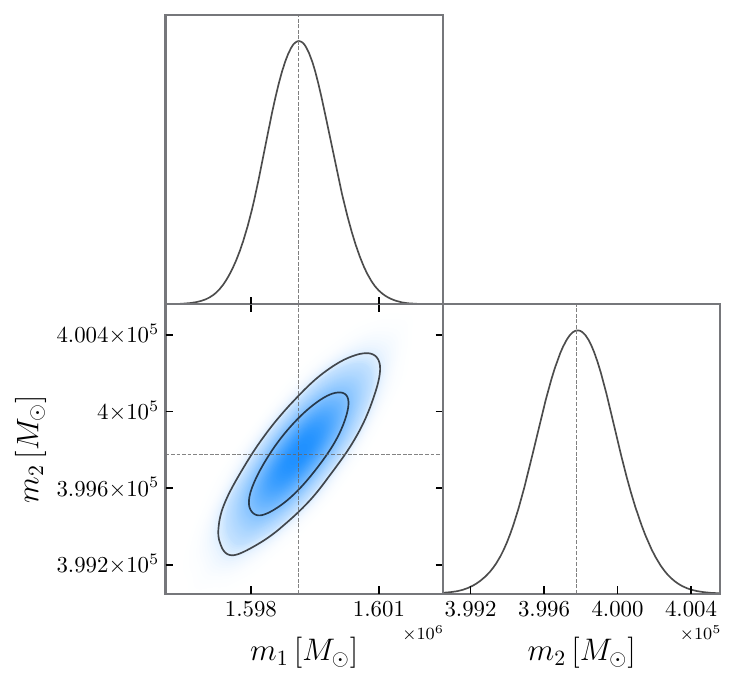}
     \caption{Corner plot showing the joint posteriors on the source frame component masses. The mass of the primary BH is measured to within $\mathcal{O}(10^3 M_{\odot})$ and the mass of the secondary to $\mathcal{O}(10^2 M_{\odot})$.
     }
     \label{fig:mass_constraints}
\end{figure}

\subsection{Constraining Binary Parameters and Early Warning}

\begin{figure*}[th!]
     \centering
     \includegraphics[width=\textwidth]{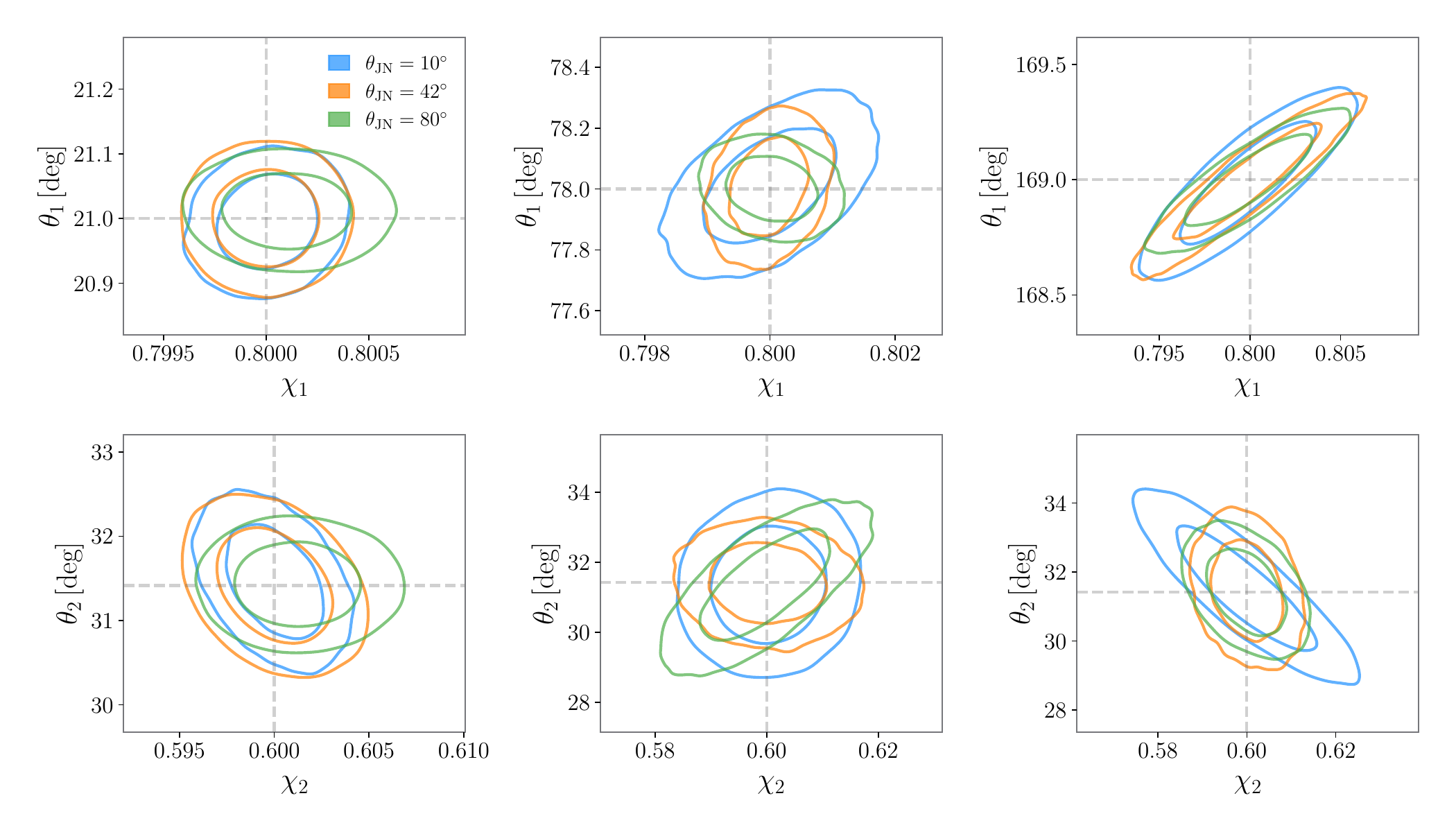}
     \caption{Constraints on the spin magnitude and orientations for the high-spin series. The top row corresponds to the primary (heavier) BH and the bottom row the secondary (lighter) BH. The left-most panels depict a configuration where the primary spin is approximately aligned with $\Lorb$, the middle panels a highly-precessing configuration, and the right-most panels where $\chi_1$ is approximately anti-aligned with $\Lorb$. For each configuration, we consider three orientations that approximately correspond to: face-on (blue), edge-on (orange), and face-off (green) binaries. 
     }
     \label{fig:spin_constraints}
\end{figure*}

\begin{figure}[th!]
     \centering
     \includegraphics[width=\columnwidth]{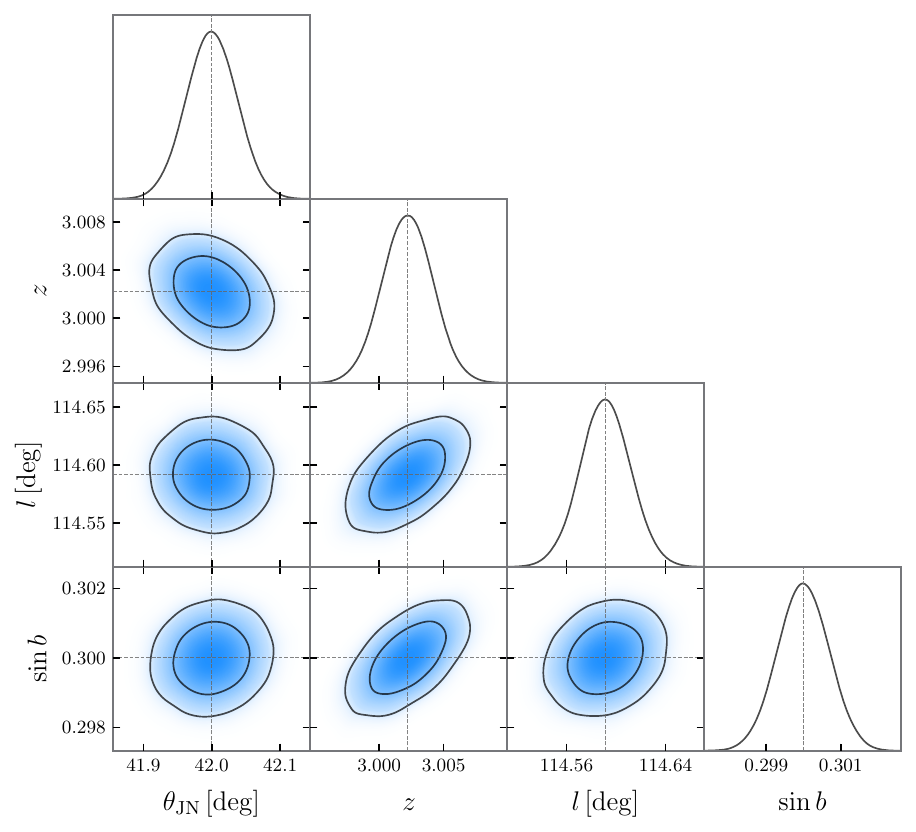}
     \caption{Corner plot showing the sky localization, redshift $z$, and the polar inclination angle of the binary $\iota$ in degrees.
     }
     \label{fig:sky_loc_constraints}
\end{figure}

\begin{figure}[th!]
     \centering
     \includegraphics[width=\columnwidth]{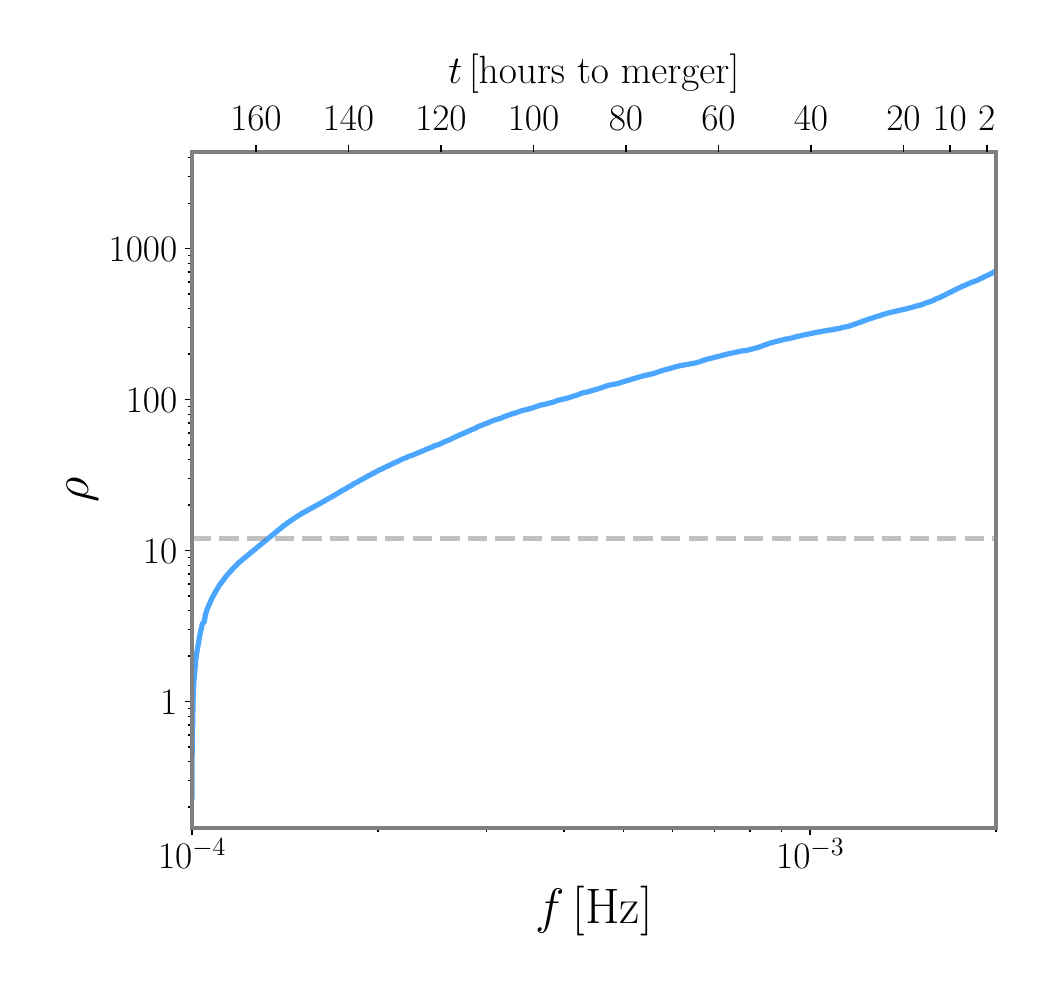}
     \caption{Growth of the signal-to-noise ratio (SNR) as a function of frequency or, equivalently, the time to merger (in hours). The horizontal dashed line denotes $\rho_{\rm det} = 12$, which we use as a proxy for the SNR at which a binary can be detected. For the canonical binary considered here, the system is visible up to $\approx 150$ hours before merger.  
     }
     \label{fig:snr}
\end{figure}

\begin{figure*}[ht!]
     \centering
     \includegraphics[width=\textwidth]{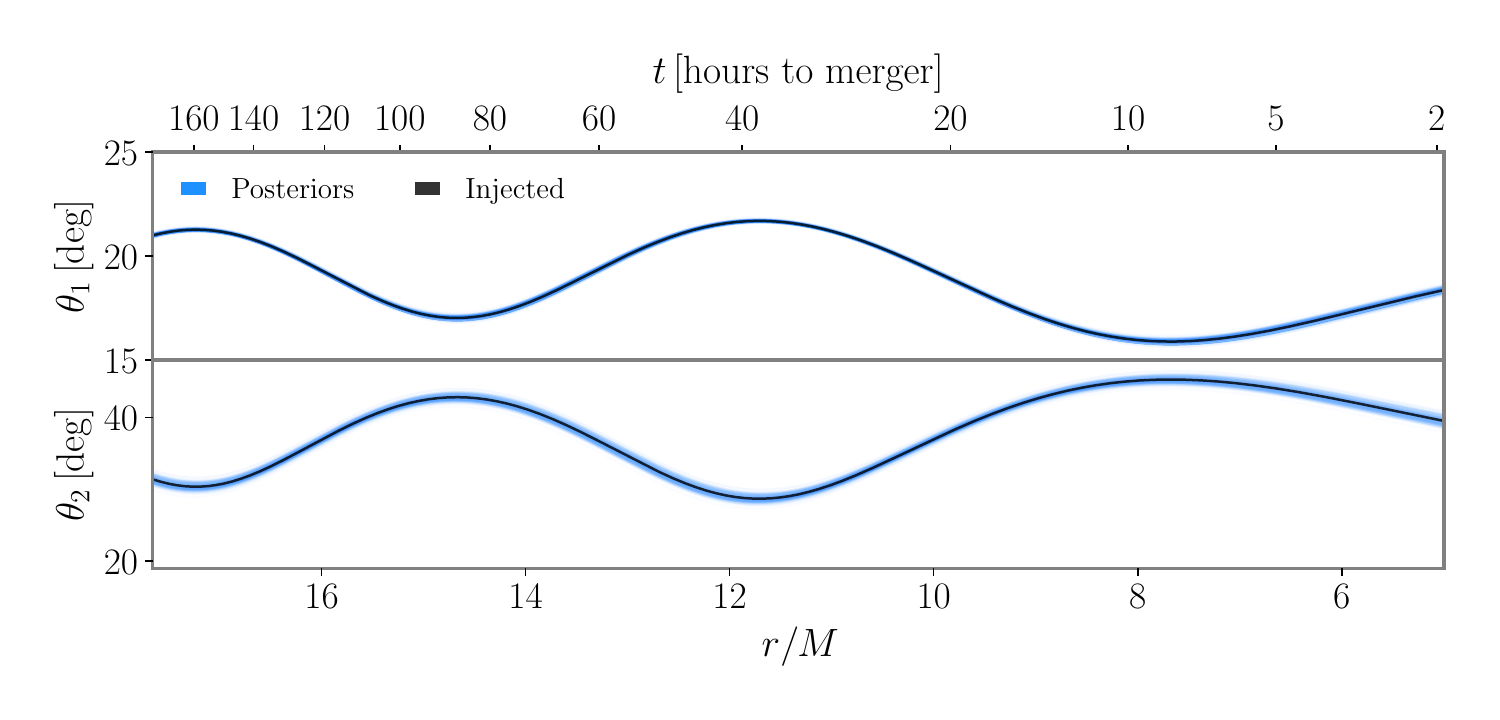}
     \\
     \includegraphics[width=\textwidth]{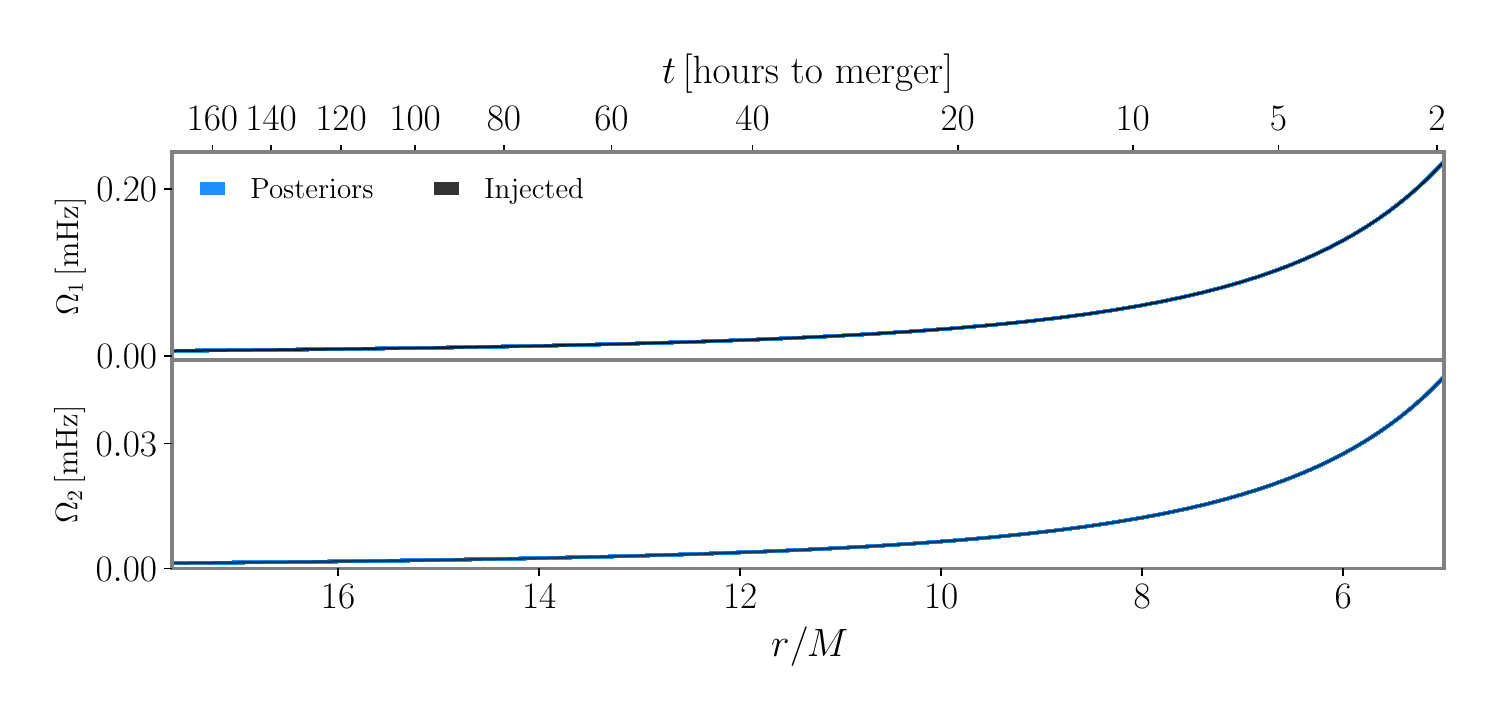}
     \caption{Tracking precession of binary HS2. Top panel: Posterior distributions on the spin evolution from the initial binary separation defined by the starting frequency of $f_{\rm low} = 10^{-4}$ Hz down to a separation of $r = 5M$, where we terminate the PN integration. On the top axis, we show the approximate time to merger in days, noting that we are considering a MBHB with a total redshifted mass of $M_{z} = 8 \times 10^6 M_{\odot}$ or, equivalently, a source-frame mass of $M_{\rm src} = 2 \times 10^6 M_{\odot}$. As noted in \cite{Pratten:2022kug}, a binary of this mass is expected to emit GWs in the LISA band for $\tau \approx 6$ days. Bottom panel: Posterior distributions on the evolution of the spin precession frequencies $\Omega_i$ in Hz as defined in Eq.~\eqref{eq:sprecfreq}. Towards the end of the inspiral, the primary BH has a precession frequency $\mathcal{O}(20 \,\rm{mHz})$ and the secondary BH $\mathcal{O}(0.5 \,\rm{mHz})$. 
     }
     \label{fig:spin_evolution}
\end{figure*}

In this study, we focus on a systematic series of investigations centred on two MBHBs that correspond to a low-spin (LS) and a high-spin (HS) configuration respectively. For each configuration, we consider three different inclination angles for the binary $\theta_{\rm JN}$ and three different spin tilt angles $\cos \theta_i = \vec{\chi}_i \cdot \Lorb$. The three spin tilts chosen correspond to a series of binaries in which the primary spin is approximately i) aligned with $\Lorb$, ii) perpendicular to $\Lorb$, and iii) anti-aligned with $\Lorb$. See Tables~\ref{tab:low_spin} and \ref{tab:high_spin} in App.~\ref{sec:binary_tables} for further details. Throughout this paper, we adopt HS2 as a canonical binary configuration, corresponding to a high-spin binary with $\chi_1 = 0.8$, $\chi_2 = 0.6$, $\iota = 42$ degrees, $\theta_1 = 21$ degrees and $\theta_2 = 31.4$ degrees. Throughout this work, we keep the redshift of the binaries the same, $z = 3$, as well as the detector-frame (i.e. redshifted) masses, corresponding to $6.4\times 10^6\,M_{\odot}$ and $1.6\times 10^5\,M_{\odot}$\footnote{We adopt the Planck18 cosmology throughout this paper \cite{Planck:2018vyg}.}. 

Even at a redshift of $z \approx 3$, the set of binaries considered here have large signal-to-noise-ratio $\rho \approx \mathcal{O}(10^3)$ resulting in excellent constraints on the BH masses and spins, irrespective of the orientation of the binary $\theta_{\rm JN}$ or the orientation of the primary BH spin $\theta_1$. As in other recent studies of MBHBs, e.g. \cite{Marsat:2020rtl,Katz:2020hku,Pratten:2022kug}, we find that the individual masses are resolved to $\Delta m_i \approx \mathcal{O}(0.1\%)$, as shown in Fig.~\ref{fig:mass_constraints}.  

For both the high- and low-spin binaries, we broadly find that the spin magnitude of the primary is measured to an error on the order of $\Delta \chi_1 \approx \mathcal{O}(0.1 \%)$, the secondary spin magnitude to $\Delta \chi_2 \approx \mathcal{O}(1 \%)$, the primary tilt to within $\Delta \theta_1 \approx \mathcal{O}(1 \%)$, and the secondary tilt to $\Delta \theta_2 \approx \mathcal{O}( 5 \%)$, as shown in Fig.~\ref{fig:spin_constraints} for the high-spin series and Fig.~\ref{fig:spin_evolution_ls} in the Appendix for the low-spin series. A full list of recovered parameters and their 90\% credible intervals can be found in Tab~\ref{tab:posteriors}. As we are able to resolve the spin magnitudes and orientations to exquisite precision, there is no need to rely on effective spin parameterizations, such as those proposed in \cite{Schmidt:2014iyl,Gerosa:2020aiw}.

Finally, we provide an assessment of the precision to which LISA can constrain the sky localization of the binary system when using an IMR waveform model that incorporates spin precession. For all binaries, the measurement accuracy for the ecliptic longitude is approximately $\Delta l \approx \mathcal{O}(0.05\%)$, while the ecliptic latitude is constrained to $\Delta b \approx \mathcal{O}(1\%)$. Furthermore, the inclination angle is determined to a precision of $\Delta \theta_{\rm JN} \approx \mathcal{O}(0.1\%)$, and the redshift to approximately $\approx \mathcal{O}(0.1\%)$. For the luminosity distance, we find that the distance to the binaries can be measured to with $\Delta d_L \approx \mathcal{O}(0.01 \rm{Gpc})$ which, assuming a Planck18 cosmology \cite{Planck:2018vyg}, translates to $\Delta z \approx \mathcal{O}(0.01)$.

These measurement accuracies correspond to a 90\% credible sky area of $\Delta \Omega_{90} \approx \mathcal{O}(0.01 \rm{deg}^2)$ and a 90\% confidence sky localization volume of $\Delta V_{90} \approx 7 \times 10^{-4} \rm{Gpc}^{3}$. In contrast to \cite{Pratten:2022kug}, which used an aligned-spin model, our findings indicate an improvement in the angular resolution of LISA due to the incorporation of spin precession, in agreement with previous studies \cite{Vecchio:2003tn,Lang:2006bsg,Lang:2007ge}. The complete sky localization posteriors are shown in Fig.~\ref{fig:sky_loc_constraints}. The accuracy to which we can constrain the sky location and distance plays a critical role in determining the multimessenger prospects, e.g see the recent detailed study in \cite{Mangiagli:2022niy} which outlines the number and properties of EM-bright signals that will be observable in both LISA and a subset of EM facilities. 

Finally, we note that the sky localization we report above is based on the analysis of the complete data from when it first enters the LISA band at $f_{\rm low} \approx 10^{-4} \rm{Hz}$ through to merger. In reality, the binary could be detectable before merger, depending on how the SNR accumulates. If a binary can be detected in sufficiently low-latency, then a preliminary skymap could be produced using partial data, e.g. adapting the methodology of \cite{Singer:2015ema}, followed by a skymap generated using the complete data. As shown in Fig.~\ref{fig:snr}, the SNR rapidly accumulates and passes a detection threshold, arbitrarily taken to be $\rho_{\det} \approx 12$, on the order of $150$ hours before merger. This would provide ample time to request a protected LISA science data period to ensure the merger is recorded based on the current lead time estimate of $\approx 2$ days~\cite{ESA:2022srd}. Note, however, that we do not make any attempt to tackle challenges pertaining to the the low-latency detection and localization of these binaries on the $\approx \mathcal{O}(\rm{hrs})$ timescale.   

\subsection{Constraining Spin Evolution}
In gas-rich environments, the binary can undergo a gas driven migration phase until the viscous timescales become smaller than the gravitational radiation-reaction timescale. At later stages of the inspiral, the binary evolution will be driven by GW emission. For the binaries considered here, the initial separation $r_{\rm i} = r(f_{\rm low})$ is within the GW-dominated phase and we can neglect disk-migration effects on the evolution. In addition, we note that if one assumes that the total angular momentum of the binary is preferentially aligned with the circumbinary disk, then the spin tilt angles $\theta_i$ are equivalent to those inherited from the gas-driven migration, see recent work in \cite{Combi:2021dks,Liao:2022niz,Steinle:2022jmc,Avara:2023ztw,Dittmann:2023dss}.

We therefore evolve the binary from an initial separation $r_{\rm i}$ to a final separation of $r_{\rm f} = 5 M$ using an orbit-averaged post-Newtonian evolution \cite{Gerosa:2016sys,Gerosa:2023xsx}. The spin tilt angles of each BH are constrained to 1 degree or better which, when combined with the stringent mass constraints at the 0.01\% level, allows for a precise determination of the evolution of the individual BH spins. LISA will have the capability to accurately ascertain the magnitudes and orientations of the spins at a comparable level of precision, down to approximately 1 hour before the merger, as depicted in Fig.~\ref{fig:spin_evolution}. As we shall see in the next section, detailed knowledge of the spin orientations at merger is vital if we are to accurately infer the recoil velocity and orientation. 
Finally, in the bottom panel of Fig.~\ref{fig:spin_evolution} we show the contributions of each spin $\boldsymbol{\Omega}_{i}$ to the precession frequency $\boldsymbol{\Omega}_{L}$, as defined in Eqns. \eqref{eq:sprecfreq} - \eqref{eq:sprecfreq2}. Precession has been measured in binary pulsars with typical precession frequencies being on the order of $10^{-10}$ Hz, e.g. \cite{Kramer:1998id,Breton:2008xy,Manchester:2010dh,Perera:2010pre,Fonseca:2014pre}. For the MBHs considered here, the precession frequencies fall in the mHz regime.  

\subsection{Constraining Remnant Properties}
As is well known, GWs carry energy and angular momentum away from the system, driving the inspiral of the binary. The orbit shrinks until the two BHs eventually plunge and form a single, perturbed BH. The emission of GWs transfers linear momentum away from the binary, causing a displacement to its center of mass in the opposite direction \cite{Bonnor:1961rec,Peres:1962rec,Bekenstein:1973rec,Fitchett:1983rec}. However, the loss of linear momentum through the merger is asymmetric and imparts a net linear velocity, i.e. a \textit{recoil} or a \textit{kick}, onto the remnant BH \cite{Gonzalez:2006md,Campanelli:2007cga,Gonzalez:2007hi,Herrmann:2007ac,Lousto:2011kp,Varma:2020nbm,Varma:2022pld}. The magnitude and orientation of the recoil has important astrophysical implications, with the first evidence for large recoil velocities being recently reported \cite{Varma:2022pld}. This mechanism can displace or even eject BHs from the galactic center \cite{Merritt:2004xa}, potentially influencing the evolution of the MBH and its host galaxy \cite{Komossa:2008as,Blecha:2010dq}. Recoils approaching $10^3 \, \rm{km} \, \rm{s}^{-1}$ could have a velocity that exceeds the escape velocity of the galactic host \cite{Redmount:1989rec,Merritt:2004xa}, suppressing the fraction of galaxies that host a central MBH \cite{Volonteri:2010rec,Gerosa:2014gja}, which has implications for the estimated event rates in LISA \cite{Sesana:2007zk,Sesana:2008ur,LISA:2022yao}. 

If a MBH is surrounded by a circumbinary disk, the perturbations to the gas induced by a recoiling BH can potentially give rise to an EM counterpart \cite{Milosavljevic:2004cg,ONeill:2008sat,Schnittman:2008ez,Lippai:2008fx,Megevand:2009rec,Rossi:2010emb,Komossa:2012cy,Piro:2022zos}. The detectability of the EM counterpart is expected to depend significantly on the kick velocity of the recoiling BH and, potentially, to the orientation relative to the circumbinary disk. For the signal to be observable, a large recoil velocity on the order of $\mathcal{O}(10^3 \, \rm{km} \, \rm{s}^{-1})$ would be necessary. Kick velocities closer to $\mathcal{O}(10^2 \, \rm{km} \, \rm{s}^{-1})$ are unlikely to produce an observable EM counterpart, based on current theoretical models \cite{Komossa:2012cy}. The EM emission spectra will be highly reliant on the specific manner in which the disk undergoes shocks \cite{Piro:2022zos}. Joint GW-EM inference will allow us to build a comprehensive understanding of the astrophysical processes at play in MBHB mergers. LISA will be able to accurately measure the masses and spins of the BHs as it approaches the merger, providing vital information on the geometry of the binary. 

\begin{figure*}
     \centering
     \begin{tabular}{lr}
     \includegraphics[width=0.45\textwidth]{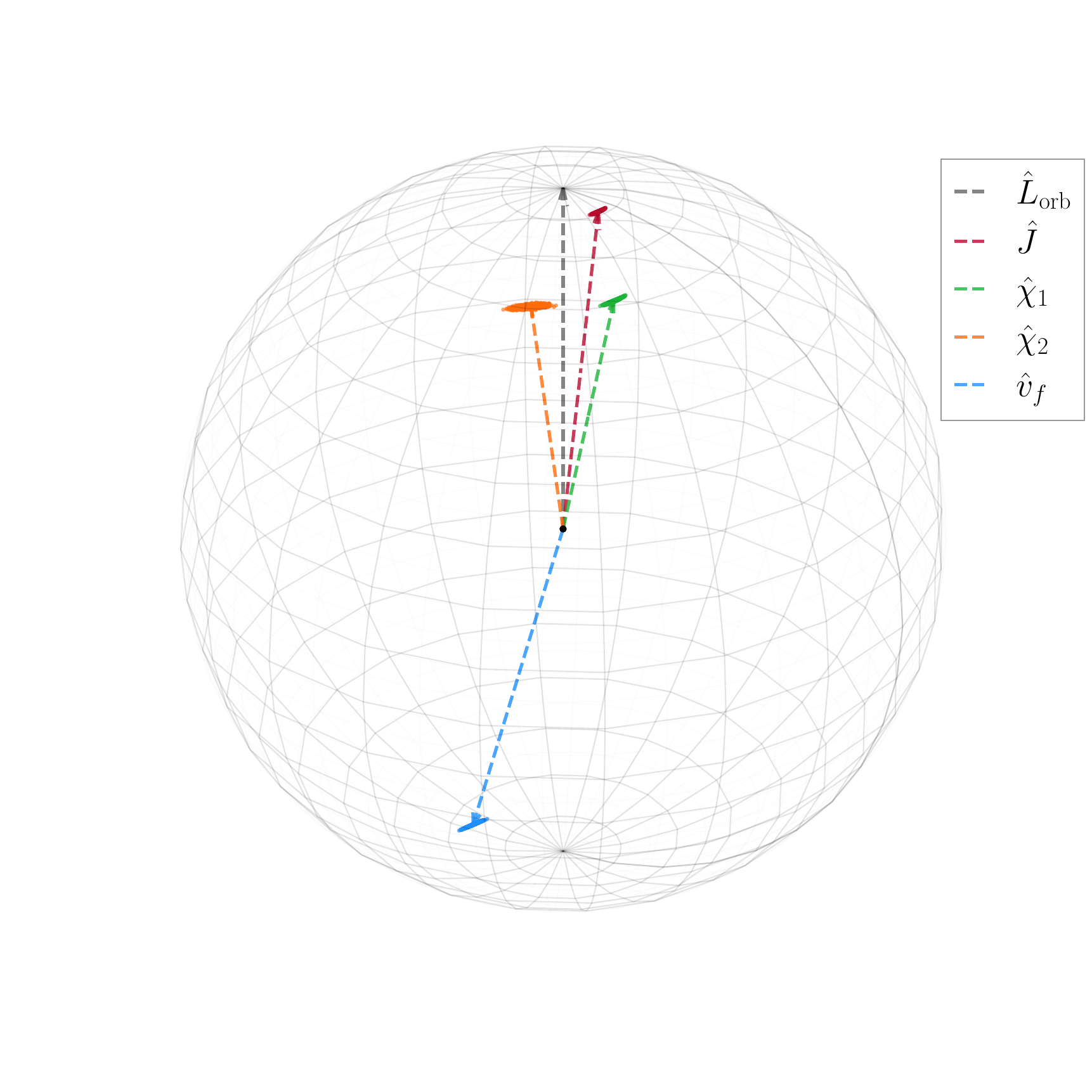}
     &
     \includegraphics[width=0.55\textwidth]{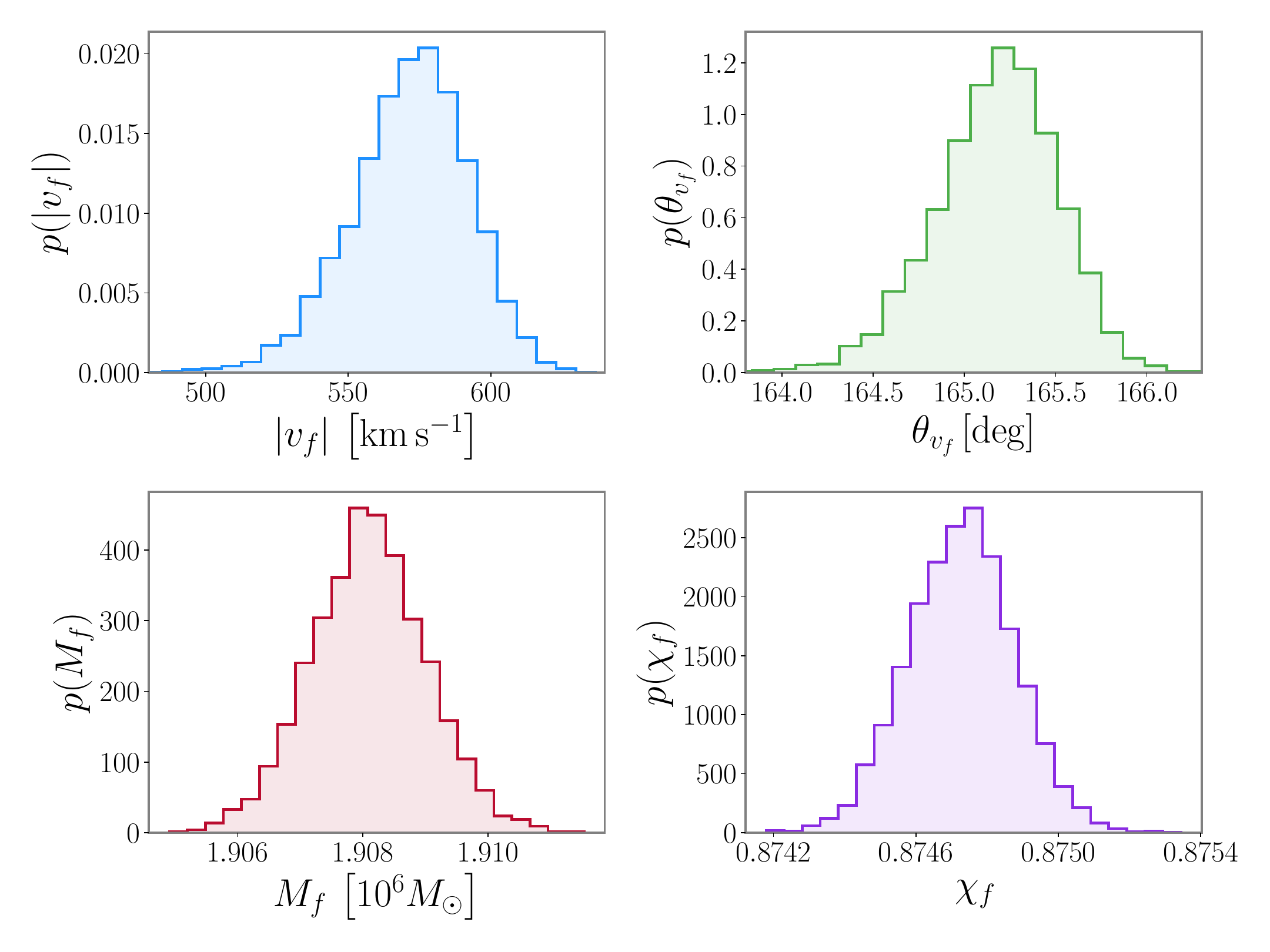}
     \\
     \multicolumn{2}{c}{\includegraphics[width=\textwidth]{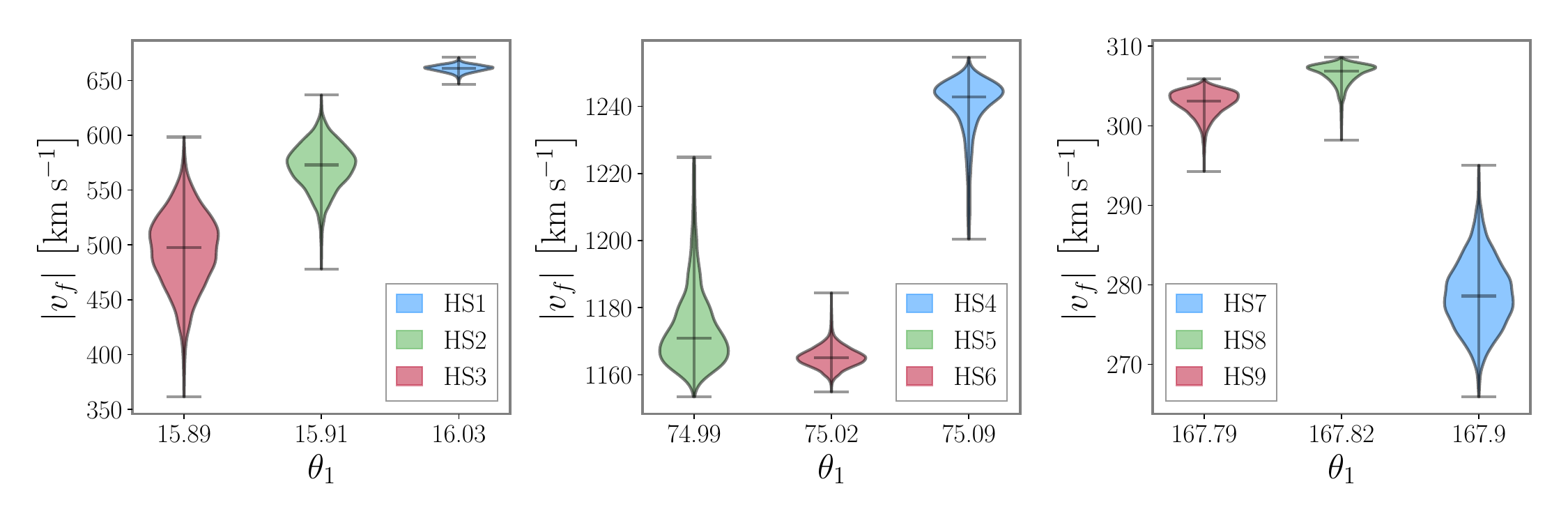}}
     \end{tabular}
    \caption{The top panels refer to the canonical ``HS2'' binary configuration. Top left: Posterior distributions for normalized total angular momentum and the dimensionless spin vectors of each BH. All quantities are defined in the orbital angular momentum frame $\hat{L} = \hat{z}$ at $t=-100M$ \cite{Varma:2018aht}. Top right: Posterior distributions for the magnitude of the kick velocity $v_f$, the polar angle of the kick orientation, the remnant source-frame BH mass $M_f$, and the dimensionless BH spin $\chi_f$. The bottom panels show the effect of different geometrical spin configurations for the ``high spin binary" on the posterior distributions for the magnitude of the kick velocity $v_f$. Here $\theta_1$ is the polar angle of the primary spin defined at a time $t = -100M$, as used to estimate the recoil velocity. 
     }
     \label{fig:recoil}
\end{figure*}

In order to understand the accuracy with which we can infer the remnant properties, we evolve the spins from $f_{\rm low}$ through to the merger, as above, and use the NR calibrated fits of \cite{Varma:2018aht} to estimate the mass $M_f$ of the remnant BH, the magnitude of the kick velocity $|v_f|$, and the orientation of the kick velocity $\theta_{v_f}$. For the canonical binary we have focused on here, we find a moderate kick velocity of $|v_f| \approx 575 \,\rm{km} \, \rm{s}^{-1}$ with an orientation of $\theta_{v_{f}} \approx 165$ deg, as shown in the top right panel of Fig.~\ref{fig:recoil}. The top left panel of Fig.~\ref{fig:recoil} highlight the precision to which LISA will be able to resolve the binary geometry in terms of the orientations of $\hat{J}, \hat{\chi}_1, \hat{\chi}_2$, and $\hat{v}_f$. As discussed in \cite{Varma:2018aht}, these quantities are defined in an instantaneous frame in which the total orbital angular momentum $\hat{L}_{\rm orb}$ points along $\hat{z}$ at a reference time of $t=-100M$. Whilst a kick velocity of $v_f \approx 575 \, \rm{km} \, \rm{s}^{-1}$ is likely to be too small to result in an EM-bright signal \cite{Piro:2022zos}, the detection and accurate characterisation of the BH recoil through GW observations alone will allow us to place important constraints on the underlying astrophysical processes. See also the discussion in \cite{Gerosa:2015xya}, which explores differential misalignment and the implications for the kick velocity. As discussed in \cite{Gerosa:2015xya}, kick velocities on the order of $\approx 1500 \rm{km} \, \rm{s}^{-1}$ can be produced in highly spinning binaries, with larger kick velocities only being anticipated for comparable mass BHs due to the dynamical alignment of the BH spins. The bottom panel of Fig.~\ref{fig:recoil} shows the posterior distributions on the kick velocity as a function of the tilt of the primary spin at $t = -100M$ for all configurations in the HS series. For the binaries in which the primary spin is essentially orthogonal to $\hat{L}$, the approaches a maximum on the order of $|v_f| \approx 1200 \, \rm{km} \, \rm{s}^{-1}$. 

\section{Conclusions and Future Work}
In this work, we have revisited the Bayesian parameter estimation of MBHBs in LISA using a state-of-the-art inspiral-merger-ringdown waveform model that incorporates both precession and higher multipoles \cite{Pratten:2020ceb,Pratten:2020fqn,Garcia-Quiros:2020qpx}. A key focus was to investigate LISA's capability to accurately constrain the masses and spins of MBHs using the complete GW signal. Given the precision to which LISA can resolve the individual spin magnitudes and orientations, we highlighted how LISA will be able to accurately track the evolution of the BH spins through to the merger. An important consequence is that LISA will be able to accurately predict the magnitude and orientation of the recoil velocity, a key observable in exploring potential EM-bright counterparts to MBH mergers.

In addition to the intrinsic parameters, we also demonstrated that, irrespective of the binary geometry, LISA will be able to constrain the sky location and orientation of the binary to sub-percent level, event out to a redshift of $z \approx 3$, at least for the subset of binaries considered here. A detailed study exploring a larger population of MBHBs is underway to explore the broader parameter space and the concomitant implications for joint GW-EM analyses. 

The detection rate of MBHBs is anticipated to be sensitive to the astrophysical distribution of spin magnitudes and orientations, especially for heavy binaries with $M > 10^7 M_{\odot}$. Detecting and inferring the properties of these binaries will play a critical role in understanding the astrophysical origin and evolution of MBHs throughout the Universe. This also has important implications in exploring the complex interplay between the MBH population observed by LISA and the stochastic GW background observed by PTAs, for which significant evidence has now been observed \cite{NANOGrav:2023gor,NANOGrav:2023hfp,Antoniadis:2023lym,Antoniadis:2023ott,Antoniadis:2023xlr,Reardon:2023gzh,Xu:2023wog}. In a recent paper, \cite{Steinle:2023vxs} demonstrated the implications of observing this PTA background on the anticipated detection rate of MBHBs in LISA. Although \cite{Steinle:2023vxs} focused on a systematic suite of non-spinning binaries, the framework can be expanded to incorporate MBHBs with generically oriented spins, as explored here.  

Detailed modeling of spin-precession plays a crucial role in resolving parameter degeneracies and enhancing the precision with which we can measure binary parameters \cite{Vecchio:2003tn,Lang:2006bsg,Klein:2009gza,Chatziioannou:2014coa,Pratten:2020igi}. It also plays a vital role in exploring MBH formation channels and constraining the astrophysics that drives their mergers. However, an important issue that remains unaddressed here is that of waveform systematics. The stringent accuracy requirements for future waveform models, necessary to reach this level of precision, pose immense challenges for source modelling. Recent developments, such as those in \cite{Khalil:2023kep,Ramos-Buades:2023ehm,Pompili:2023tna,vandeMeent:2023ols,Yu:2023lml,Hamilton:2021pkf,Wardell:2021fyy}, have shown progress in enhancing the accuracy of the current generation of waveform models. Yet, achieving the accuracy \textit{and} data analysis demands of LISA still remains a significant endeavor. In future work, we will explore the impact of waveform systematics on our ability to constrain the astrophysics behind MBH mergers. Moreover, we note that the GW signal can be influenced by environmental effects, such as dynamical friction or torquing from angular momentum exchange between accretion discs and the binary \cite{Barausse:2007dy,Kocsis:2011dr,Barausse:2014tra,Speri:2022upm}. Detailed phenomenological modeling of environmental effects opens a promising avenue to discern the astrophysical processes at play in galactic nuclei but could introduce systematic biases into our analysis \cite{Zwick:2022dih}.

The results presented here are a first step towards integrating waveform models with precession and higher multipoles into a global fit pipeline. Notably, \cite{Littenberg:2023xpl} has demonstrated the application of aligned-spin phenomenological waveform models to address the MBHB aspect of the global fit. In our case, we have employed IMRPhenomXPHM to perform Bayesian inference on individual MBHBs. Finally, we note that IMRPhenomXPHM can be effectively combined with the heterodyned likelihood \cite{Cornish:2021lje}, even when including precession and higher multipoles \cite{Leslie:2021ssu}. In forthcoming work, we will build upon the findings presented here to demonstrate how MBHBs can effectively constrain the underlying astrophysical population \cite{Pratten:2023pop} and facilitate tests of General Relativity \cite{Pratten:2023tgr}. 

\section*{Acknowledgments}
We thank Cl\'ement Bonnerot and Jean-Baptiste Bayle for useful comments on the manuscript.
GP gratefully acknowledges support from a Royal Society University Research Fellowship URF{\textbackslash}R1{\textbackslash}221500 and RF{\textbackslash}ERE{\textbackslash}221015. PS and GP acknowledge support from STFC grant ST/V005677/1. HM and AV acknowledge the support of the UK Space Agency Grant No. ST/V002813/1 and ST/X002071/1. AV acknowledges the support of the Royal Society and Wolfson Foundation. Computations were performed on the University of Birmingham’s high performance computing service BlueBEAR and the Bondi HPC Cluster at the Birmingham Institute for Gravitational Wave Astronomy. Parts of this analysis made use of \texttt{numpy} \cite{numpy}, \texttt{matplotlib} \cite{matplotlib}, \texttt{scipy} \cite{scipy}, \texttt{numpy} \cite{numpy}, \texttt{GetDist} \cite{Lewis:2019xzd}, \texttt{precession} \cite{Gerosa:2016sys,Gerosa:2023xsx}, and \texttt{dynesty} \cite{Koposov:2023dyn}.

\appendix
\section{Binaries}
\label{sec:binary_tables}

\begin{table}[ht!]
\def\arraystretch{1.5}
\centering
\begin{tabular}{l|cc|cc|ccc|c|ccc}
\toprule
ID & $m_{1}$ & $m_{2}$ & $\chi_1$ & $\chi_2$ & $\theta_1$ & $\theta_2$ & $\phi_{12}$ & $z$ & $\iota$ & $b$ & $l$ \\
 & \multicolumn{2}{c|}{$\left[ 10^6 M_{\odot} \right]$}
&
&
& \multicolumn{3}{c|}{$\left[ \rm deg \right]$}
&
& \multicolumn{3}{c}{$\left[ \rm deg \right]$} \\
\hline
\hline
LS1 &
$1.6$&
$0.4$&
$0.2$&
$0.4$&
$21.0$&
$31.4$&
$34.4$&
$3.0$&
$10.0$&
$17.5$&
$114.6$
\\
LS2 &
$-$&
$-$&
$-$&
$-$&
$-$&
$-$&
$-$&
$-$&
$42.0$&
$-$&
$-$
\\
LS3 &
$-$&
$-$&
$-$&
$-$&
$-$&
$-$&
$-$&
$-$&
$80.0$&
$-$&
$-$
\\
\hline
LS4 & 
$1.6$&
$0.4$&
$0.2$&
$0.4$&
$78.0$&
$31.4$&
$34.4$&
$3.0$&
$10.0$&
$17.5$&
$114.6$
\\
LS5 & 
$-$&
$-$&
$-$&
$-$&
$-$&
$-$&
$-$&
$-$&
$42.0$&
$-$&
$-$
\\
LS6 & 
$-$&
$-$&
$-$&
$-$&
$-$&
$-$&
$-$&
$-$&
$80.0$&
$-$&
$-$
\\
\hline 
LS7 &
$1.6$&
$0.4$&
$0.2$&
$0.4$&
$169.0$&
$31.4$&
$34.4$&
$3.0$&
$10.0$&
$17.5$&
$114.6$
\\
LS8 & 
$-$&
$-$&
$-$&
$-$&
$-$&
$-$&
$-$&
$-$&
$42.0$&
$-$&
$-$
\\
LS9 & 
$-$&
$-$&
$-$&
$-$&
$-$&
$-$&
$-$&
$-$&
$80.0$&
$-$&
$-$
\\
\hline 
\end{tabular}
\caption{Summary table for the \textit{low spin} (LS) binaries, characterized by $\chi_1 = 0.2$ and $\chi_2 = 0.4$. We consider three configurations, corresponding to $\theta_1 = \lbrace 21{\degree}, 78{\degree}, 169{\degree} \rbrace$ degrees respectively. For each configuration, we consider three inclination angles $\iota = \lbrace 10{\degree}, 42{\degree}, 80{\degree} \rbrace$ to gauge the impact of the binary orientation on our ability to resolve the binary parameters. Here $10$ degrees corresponds to a near face-on configuration and $80$ degrees to an approximately edge-on configuration. The masses correspond to the source-frame values and a dashed line denotes that the value is equivalent to the row above. 
}
\label{tab:low_spin}
\end{table}

\begin{table} 
\def\arraystretch{1.5}
\centering
\begin{tabular}{l|cc|cc|ccc|c|ccc}
\toprule
ID & $m_{1}$ & $m_{2}$ & $\chi_1$ & $\chi_2$ & $\theta_1$ & $\theta_2$ & $\phi_{12}$ & $z$ & $\iota$ & $b$ & $l$ \\
 & \multicolumn{2}{c|}{$\left[ 10^6 M_{\odot} \right]$}
&
&
& \multicolumn{3}{c|}{$\left[ \rm deg \right]$}
&
& \multicolumn{3}{c}{$\left[ \rm deg \right]$} \\
\hline
\hline
HS1 &
$1.6$&
$0.4$&
$0.8$&
$0.6$&
$21.0$&
$31.4$&
$34.4$&
$3.0$&
$10.0$&
$17.5$&
$114.6$
\\
HS2 &
$-$&
$-$&
$-$&
$-$&
$-$&
$-$&
$-$&
$-$&
$42.0$&
$-$&
$-$
\\
HS3 &
$-$&
$-$&
$-$&
$-$&
$-$&
$-$&
$-$&
$-$&
$80.0$&
$-$&
$-$
\\
\hline
HS4 & 
$1.6$&
$0.4$&
$0.8$&
$0.6$&
$78.0$&
$31.4$&
$34.4$&
$3.0$&
$10.0$&
$17.5$&
$114.6$
\\
HS5 & 
$-$&
$-$&
$-$&
$-$&
$-$&
$-$&
$-$&
$-$&
$42.0$&
$-$&
$-$
\\
HS6 & 
$-$&
$-$&
$-$&
$-$&
$-$&
$-$&
$-$&
$-$&
$80.0$&
$-$&
$-$
\\
\hline 
HS7 &
$1.6$&
$0.4$&
$0.8$&
$0.6$&
$169.0$&
$31.4$&
$34.4$&
$3.0$&
$10.0$&
$17.5$&
$114.6$
\\
HS8 & 
$-$&
$-$&
$-$&
$-$&
$-$&
$-$&
$-$&
$-$&
$42.0$&
$-$&
$-$
\\
HS9 & 
$-$&
$-$&
$-$&
$-$&
$-$&
$-$&
$-$&
$-$&
$80.0$&
$-$&
$-$
\\
\hline 
\end{tabular}
\caption{As in Tab.~\ref{tab:low_spin}, but now for the \textit{high spin} (HS) binaries, where $\chi_1 = 0.8$ and $\chi_2 = 0.6$. 
}
\label{tab:high_spin}
\end{table}

\section{Summary of Bayesian Inference}
\label{sec:app_table}
Here we list the exact configurations used for the low-spin (LS) and high-spin (HS) binaries. For each set of binaries we fix the sky-location, distance, component masses, spin magnitudes, and spin tilt of the secondary BH. We then systematically vary the spin tilt of the primary BH from an approximately aligned configuration through to an approximately anti-aligned configuration. Spins that are misaligned with the orbital angular momentum give rise to relativistic spin couplings that drive the precession of the orbital plane and the spins themselves \cite{Apostolatos:1994mx,Kidder:1995zr}. By tilting the primary spin we are effectively changing the amount of precession in the binary. In addition to varying the primary spin magnitude, we also explore three binary orientations for each tilt configuration to explore the impact of observing the binary from a near face-on configuration through to a near edge-on configuration. Due to the fact that the binaries undergo multiple precession cycles after entering the observable LISA band, we find that the orientation has little impact on our ability to constrain the BH spins. 

\begin{table*}[ht!]
\caption{
Summary of the 90\% credible intervals on the recovered parameters for all injections considered in this paper. Here HS denotes the high-spin series and LS the low-spin series. The component masses $m_i$ correspond to the source-frame values, $\chi_i$ the dimensionless spin magnitudes, $\theta_i$ the spin-tilt angles, and $\phi_{12}$ the difference in the azimuthal spin angles between the two BH spins. 
}
\label{tab:posteriors}
\setlength{\tabcolsep}{1pt}
\vspace{0.1cm}
\def\arraystretch{1.6}
\centering
\begin{tabular}{c|cc|ccccc|cc|cc}
\toprule
\hline
ID & $m_{1}$ & $m_{2}$ & $\chi_1$ & $\chi_2$ & $\theta_1$ & $\theta_2$ & $\phi_{12}$ & $\theta_{\rm JN}$ & $z$ & $\sin b$ & $l$ \\
& $\left[10^6 M_{\odot} \right]$ & $\left[10^6 M_{\odot} \right]$ & $-$ & $-$ & $\left[ \rm{deg} \right]$ & $\left[ \rm{deg} \right]$ & $\left[ \rm{deg} \right]$ & $\left[ \rm{deg} \right]$ & $-$ & $-$  & $\left[ \rm{deg} \right]$
\\
\hline 
\hline
HS1  &  
${1.599}^{+0.001}_{-0.001}$  &  
${0.400}^{+0.0003}_{-0.0003}$  &  
${0.800}^{+0.0003}_{-0.0003}$  &  
${0.600}^{+0.003}_{-0.003}$  &  
${21.0}^{+0.1}_{-0.1}$  &  
${31.5}^{+0.7}_{-0.7}$  &  
${34.5}^{+1.3}_{-1.3}$  &  
${10.00}^{+0.05}_{-0.05}$  &  
${3.002}^{+0.003}_{-0.003}$  &  
${0.300}^{+0.001}_{-0.001}$  &  
${114.59}^{+0.06}_{-0.07}$
\\
HS2  &  
${1.599}^{+0.001}_{-0.001}$  &  
${0.400}^{+0.0004}_{-0.0003}$  &  
${0.800}^{+0.0003}_{-0.0003}$  &  
${0.600}^{+0.003}_{-0.003}$  &  
${21.0}^{+0.1}_{-0.1}$  &  
${31.4}^{+0.7}_{-0.7}$  &  
${34.4}^{+1.5}_{-1.5}$  &  
${42.00}^{+0.06}_{-0.06}$  &  
${3.002}^{+0.003}_{-0.003}$  &  
${0.300}^{+0.001}_{-0.001}$  &  
${114.59}^{+0.03}_{-0.03}$
\\
HS3  &  
${1.599}^{+0.002}_{-0.002}$  &  
${0.400}^{+0.0005}_{-0.0005}$  &  
${0.800}^{+0.0004}_{-0.0003}$  &  
${0.601}^{+0.004}_{-0.004}$  &  
${21.0}^{+0.1}_{-0.1}$  &  
${31.4}^{+0.5}_{-0.5}$  &  
${34.5}^{+1.2}_{-1.2}$  &  
${80.00}^{+0.03}_{-0.03}$  &  
${3.003}^{+0.005}_{-0.005}$  &  
${0.300}^{+0.002}_{-0.002}$  &  
${114.59}^{+0.02}_{-0.02}$
\\
HS4  &  
${1.599}^{+0.002}_{-0.002}$  &  
${0.400}^{+0.0005}_{-0.0006}$  &  
${0.800}^{+0.0012}_{-0.0012}$  &  
${0.600}^{+0.011}_{-0.011}$  &  
${78.0}^{+0.2}_{-0.2}$  &  
${31.4}^{+1.8}_{-1.8}$  &  
${34.2}^{+2.5}_{-2.4}$  &  
${10.00}^{+0.08}_{-0.08}$  &  
${3.002}^{+0.005}_{-0.005}$  &  
${0.300}^{+0.002}_{-0.002}$  &  
${114.59}^{+0.09}_{-0.09}$
\\
HS5  &  
${1.599}^{+0.002}_{-0.002}$  &  
${0.400}^{+0.0006}_{-0.0006}$  &  
${0.800}^{+0.0007}_{-0.0007}$  &  
${0.600}^{+0.012}_{-0.011}$  &  
${78.0}^{+0.2}_{-0.2}$  &  
${31.4}^{+1.2}_{-1.3}$  &  
${34.2}^{+2.4}_{-2.3}$  &  
${42.00}^{+0.04}_{-0.04}$  &  
${3.002}^{+0.006}_{-0.006}$  &  
${0.300}^{+0.002}_{-0.002}$  &  
${114.59}^{+0.03}_{-0.03}$
\\
HS6  &  
${1.599}^{+0.002}_{-0.002}$  &  
${0.400}^{+0.0005}_{-0.0005}$  &  
${0.800}^{+0.0008}_{-0.0008}$  &  
${0.600}^{+0.013}_{-0.013}$  &  
${78.0}^{+0.1}_{-0.1}$  &  
${31.3}^{+1.7}_{-1.7}$  &  
${34.3}^{+3.1}_{-3.3}$  &  
${80.00}^{+0.05}_{-0.05}$  &  
${3.002}^{+0.004}_{-0.004}$  &  
${0.300}^{+0.002}_{-0.002}$  &  
${114.59}^{+0.02}_{-0.02}$
\\
HS7  &  
${1.599}^{+0.005}_{-0.005}$  &  
${0.400}^{+0.0012}_{-0.0012}$  &  
${0.800}^{+0.0040}_{-0.0041}$  &  
${0.599}^{+0.017}_{-0.017}$  &  
${169.0}^{+0.3}_{-0.3}$  &  
${31.5}^{+2.0}_{-1.9}$  &  
${34.6}^{+7.1}_{-7.0}$  &  
${10.00}^{+0.33}_{-0.34}$  &  
${3.002}^{+0.012}_{-0.012}$  &  
${0.300}^{+0.005}_{-0.005}$  &  
${114.59}^{+0.26}_{-0.25}$
\\
HS8  &  
${1.599}^{+0.004}_{-0.004}$  &  
${0.400}^{+0.0011}_{-0.0011}$  &  
${0.800}^{+0.0043}_{-0.0044}$  &  
${0.600}^{+0.009}_{-0.009}$  &  
${169.0}^{+0.3}_{-0.3}$  &  
${31.5}^{+1.6}_{-1.6}$  &  
${34.5}^{+5.5}_{-5.5}$  &  
${42.00}^{+0.11}_{-0.11}$  &  
${3.002}^{+0.011}_{-0.011}$  &  
${0.300}^{+0.005}_{-0.005}$  &  
${114.59}^{+0.07}_{-0.07}$
\\
HS9  &  
${1.599}^{+0.004}_{-0.004}$  &  
${0.400}^{+0.0011}_{-0.0011}$  &  
${0.800}^{+0.0038}_{-0.0037}$  &  
${0.600}^{+0.010}_{-0.010}$  &  
${169.0}^{+0.2}_{-0.2}$  &  
${31.5}^{+1.3}_{-1.3}$  &  
${34.4}^{+4.9}_{-4.9}$  &  
${80.00}^{+0.11}_{-0.11}$  &  
${3.002}^{+0.010}_{-0.010}$  &  
${0.300}^{+0.003}_{-0.003}$  &  
${114.59}^{+0.04}_{-0.04}$
\\
\hline 
LS1  &  
${1.599}^{+0.005}_{-0.005}$  &  
${0.400}^{+0.001}_{-0.001}$  &  
${0.200}^{+0.004}_{-0.004}$  &  
${0.400}^{+0.014}_{-0.014}$  &  
${21.0}^{+0.5}_{-0.5}$  &  
${31.2}^{+3.8}_{-3.8}$  &  
${34.2}^{+5.6}_{-6.0}$  &  
${10.00}^{+0.08}_{-0.08}$  &  
${3.00}^{+0.01}_{-0.01}$  &  
${0.300}^{+0.003}_{-0.003}$  &  
${114.59}^{+0.18}_{-0.18}$
\\
LS2  &  
${1.599}^{+0.003}_{-0.003}$  &  
${0.400}^{+0.001}_{-0.001}$  &  
${0.200}^{+0.002}_{-0.002}$  &  
${0.400}^{+0.012}_{-0.012}$  &  
${21.0}^{+0.6}_{-0.6}$  &  
${31.7}^{+2.7}_{-2.7}$  &  
${34.3}^{+7.5}_{-8.0}$  &  
${41.99}^{+0.13}_{-0.13}$  &  
${3.00}^{+0.01}_{-0.01}$  &  
${0.300}^{+0.003}_{-0.003}$  &  
${114.59}^{+0.11}_{-0.10}$
\\
LS3  &  
${1.599}^{+0.003}_{-0.003}$  &  
${0.400}^{+0.001}_{-0.001}$  &  
${0.200}^{+0.002}_{-0.002}$  &  
${0.399}^{+0.014}_{-0.014}$  &  
${21.0}^{+0.6}_{-0.6}$  &  
${31.2}^{+2.4}_{-2.3}$  &  
${34.9}^{+6.0}_{-6.1}$  &  
${80.00}^{+0.05}_{-0.05}$  &  
${3.00}^{+0.01}_{-0.01}$  &  
${0.300}^{+0.004}_{-0.004}$  &  
${114.59}^{+0.04}_{-0.04}$
\\
LS4  &  
${1.599}^{+0.004}_{-0.004}$  &  
${0.400}^{+0.001}_{-0.001}$  &  
${0.200}^{+0.002}_{-0.002}$  &  
${0.393}^{+0.024}_{-0.024}$  &  
${77.7}^{+1.1}_{-1.1}$  &  
${30.9}^{+2.8}_{-3.0}$  &  
${35.2}^{+6.6}_{-6.7}$  &  
${10.00}^{+0.07}_{-0.07}$  &  
${3.00}^{+0.01}_{-0.01}$  &  
${0.300}^{+0.003}_{-0.003}$  &  
${114.59}^{+0.18}_{-0.18}$
\\
LS5  &  
${1.599}^{+0.003}_{-0.003}$  &  
${0.400}^{+0.001}_{-0.001}$  &  
${0.200}^{+0.002}_{-0.002}$  &  
${0.398}^{+0.021}_{-0.022}$  &  
${77.8}^{+0.9}_{-0.9}$  &  
${31.3}^{+2.5}_{-2.8}$  &  
${35.7}^{+6.5}_{-6.7}$  &  
${42.00}^{+0.12}_{-0.12}$  &  
${3.00}^{+0.01}_{-0.01}$  &  
${0.300}^{+0.003}_{-0.003}$  &  
${114.59}^{+0.11}_{-0.11}$
\\
LS6  &  
${1.599}^{+0.003}_{-0.003}$  &  
${0.400}^{+0.001}_{-0.001}$  &  
${0.200}^{+0.002}_{-0.002}$  &  
${0.399}^{+0.016}_{-0.016}$  &  
${78.1}^{+0.8}_{-0.8}$  &  
${31.3}^{+2.0}_{-2.1}$  &  
${32.9}^{+6.2}_{-6.6}$  &  
${80.00}^{+0.07}_{-0.07}$  &  
${3.00}^{+0.01}_{-0.01}$  &  
${0.300}^{+0.004}_{-0.004}$  &  
${114.59}^{+0.04}_{-0.04}$
\\
LS7  &  
${1.599}^{+0.007}_{-0.006}$  &  
${0.400}^{+0.001}_{-0.001}$  &  
${0.200}^{+0.006}_{-0.007}$  &  
${0.400}^{+0.020}_{-0.024}$  &  
${169.0}^{+0.5}_{-0.5}$  &  
${31.3}^{+4.8}_{-3.8}$  &  
${34.1}^{+11.9}_{-14.8}$  &  
${10.00}^{+0.17}_{-0.17}$  &  
${3.00}^{+0.01}_{-0.01}$  &  
${0.300}^{+0.004}_{-0.004}$  &  
${114.60}^{+0.22}_{-0.22}$
\\
LS8  &  
${1.599}^{+0.004}_{-0.004}$  &  
${0.400}^{+0.001}_{-0.001}$  &  
${0.200}^{+0.004}_{-0.004}$  &  
${0.401}^{+0.015}_{-0.014}$  &  
${169.0}^{+0.6}_{-0.7}$  &  
${31.3}^{+4.1}_{-3.8}$  &  
${34.8}^{+12.6}_{-12.3}$  &  
${42.00}^{+0.16}_{-0.16}$  &  
${3.00}^{+0.01}_{-0.01}$  &  
${0.300}^{+0.003}_{-0.003}$  &  
${114.59}^{+0.13}_{-0.13}$
\\
LS9  &  
${1.599}^{+0.004}_{-0.004}$  &  
${0.400}^{+0.001}_{-0.001}$  &  
${0.200}^{+0.003}_{-0.003}$  &  
${0.400}^{+0.013}_{-0.013}$  &  
${169.0}^{+0.3}_{-0.3}$  &  
${31.4}^{+2.2}_{-2.1}$  &  
${34.7}^{+10.6}_{-11.1}$  &  
${80.00}^{+0.09}_{-0.09}$  &  
${3.00}^{+0.01}_{-0.01}$  &  
${0.300}^{+0.004}_{-0.004}$  &  
${114.59}^{+0.04}_{-0.04}$
\\
\hline
\end{tabular}
\end{table*}

\section*{Supplementary Plots}
\label{sec:app_plots}
\begin{figure*}[th!]
     \centering
     \includegraphics[width=\textwidth]{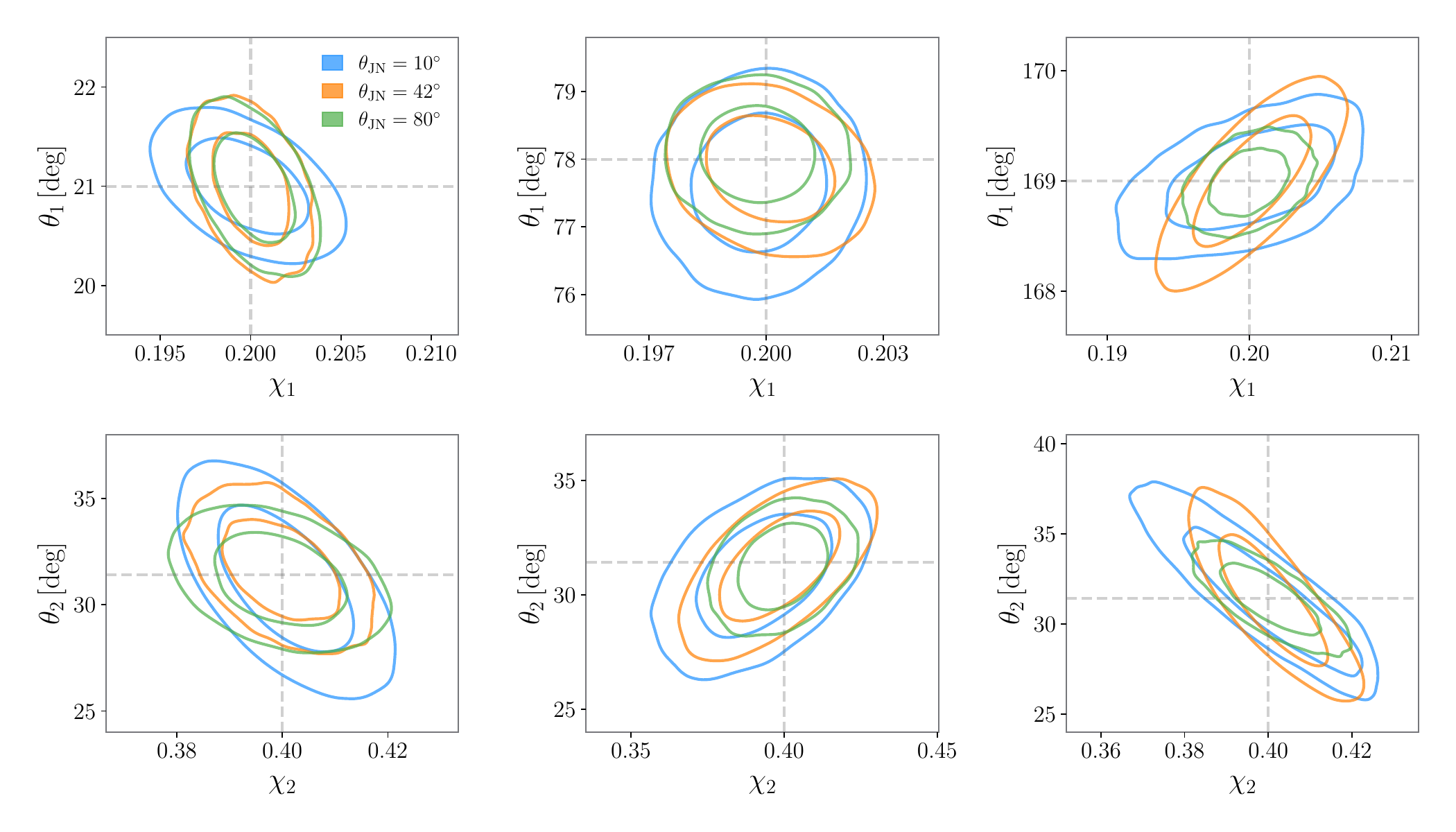}
     \caption{As per Fig.~\ref{fig:spin_constraints} but for the low-spin systematic series. 
     }
     \label{fig:spin_evolution_ls}
\end{figure*}

Figure~\ref{fig:spin_evolution_ls} is the low-spin counterpart of Fig.~\ref{fig:spin_evolution}, demonstrating that even for low spins, LISA can accurately resolve the magnitudes and orientations of BH spins to a comparable level of accuracy.

\section*{Corner Plots}
\label{sec:corner}
\begin{figure*}[th!]
     \centering
     \includegraphics[width=\textwidth]{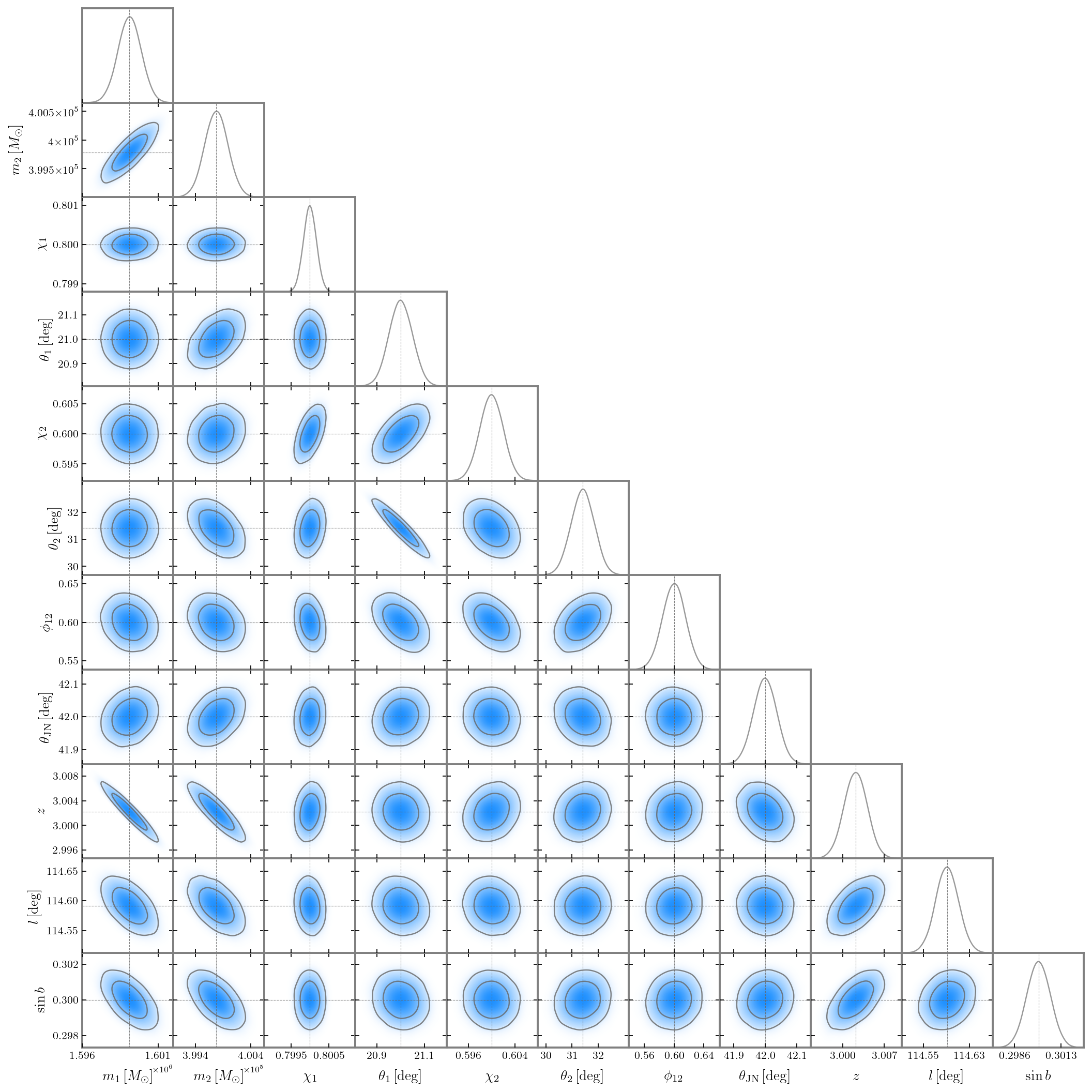}
     \caption{Full corner plot for our canonical binary. The dashed lines denote the injected values. 
     }
     \label{fig:posteriors_hs2}
\end{figure*}

The complete joint posteriors for our canonical binary configuration, HS2, are shown in Fig.~\ref{fig:posteriors_hs2}. 

\clearpage
\twocolumngrid
\bibliography{refs}

\end{document}